\documentclass[runningheads]{llncs}

\usepackage[final,year=2026]{eccv}
\usepackage{eccvabbrv}

\usepackage{graphicx}
\usepackage{booktabs}
\usepackage{siunitx}
\usepackage{makecell}

\usepackage[accsupp]{axessibility} 
\usepackage[pagebackref,colorlinks]{hyperref}
\usepackage[ruled,vlined,linesnumbered]{algorithm2e}
\DontPrintSemicolon
\usepackage{xcolor}
\usepackage{graphicx}
\usepackage{subcaption}
\usepackage{multirow}

\begin{document}

\title{Quantization as a Malicious Task: Removing Quantization-Conditioned Backdoors via Task Arithmetic} 

\titlerunning{Abbreviated paper title}

\author{
Kaihsun Yang \and
Min-Yan Tsai \and Chia-Mu Yu
}

\authorrunning{F.~Author et al.}

\institute{National Yang Ming Chiao Tung University
}

\maketitle

\begin{abstract}
Model quantization is widely adopted to reduce memory usage and inference cost when deploying deep neural networks on resource-constrained devices. However, recent studies have revealed a new security threat known as Quantization-Conditioned Backdoors (QCBs), where a model behaves normally in full precision but activates malicious behavior only after quantization. Existing defenses typically modify quantization procedures or correct activation statistics, often introducing additional computational overhead or relying on specific quantization settings. Here, we present \textsc{QVec}, a parameter-space perspective for defending against QCBs. We observe that the weight difference between a full-precision model and its quantized counterpart encodes a structured behavioral shift, which can be interpreted as a malicious task vector rather than random quantization noise. Based on this insight, \textsc{QVec} counteracts this malicious direction through controlled parameter correction prior to deployment. \textsc{QVec} requires no retraining, no trigger samples, and only a single quantization pass to estimate the parameter shift, together with a lightweight hyperparameter search. Extensive experiments across image classification benchmarks and multiple Large Language Model (LLM) attack scenarios demonstrate that \textsc{QVec} consistently suppresses backdoor activation while preserving clean performance.

\end{abstract}

\section{Introduction}
\label{sec: Introduction}
Model quantization has become a fundamental technique for deploying deep neural networks under resource constraints. By mapping full-precision weights into low-bit representations (e.g., INT8 or INT4), quantization reduces memory footprint and inference latency while maintaining competitive performance. Early work such as integer-only inference frameworks \cite{Jacob2018} enabled practical edge deployment, while Post Training Quantization (PTQ) and Quantization Aware Training (QAT) further improved robustness and accuracy \cite{Nagel2019,Banner2019,Esser2020,Choi2018}. Recent advances have extended quantization to optimizers and large-scale models \cite{Dettmers2022}, making low-bit inference increasingly common in both vision and language systems.

Meanwhile, backdoor attacks pose a serious threat to neural networks by embedding hidden malicious behaviors during training \cite{Gu2017}. Classical backdoors are triggered by specific input patterns and remain dormant otherwise. Numerous detection and purification defenses have been proposed \cite{Liu2018,Sha2022,Zeng2022,Wang2019}, but most assume the backdoor is already active in the full-precision model.

Recently, a new attack paradigm termed quantization-conditioned backdoors (QCBs) has emerged \cite{Hong2021,Ma2023TDSC,Huynh2024}. In these attacks, a model behaves benignly in full precision but activates malicious behavior only after quantization. This challenges the assumptions underlying conventional backdoor defenses and introduces a new supply-chain vulnerability. The threat has further escalated with its extension to LLMs \cite{Egashira2024}, showing that quantization itself can serve as the triggering condition.

Defending against QCBs presents unique challenges. First, the malicious behavior remains completely dormant in the full-precision model, rendering most detection-based methods ineffective. Since existing defenses analyze activation patterns, gradients, or anomalous neurons in full precision \cite{Wang2019,Liu2018}, they cannot detect QCBs prior to quantization. Second, the triggering mechanism depends on quantization-induced weight perturbations whose effects vary across quantization granularities (e.g., per-tensor vs.\ per-channel), bit-widths, and rounding schemes. As shown in \cite{Hong2021,Ma2023TDSC}, even small differences in rounding or scaling can significantly affect attack success. Finally, practical deployment often prohibits retraining or access to large clean datasets. Therefore, an effective defense should require minimal data, incur low computational overhead, and preserve model performance.

Our key observation is that QCB activation originates from the parameter shift induced by quantization. Let $W$ denote the full-precision model and $Q(W)$ its quantized counterpart, and define the quantization-induced shift as $\delta = Q(W) - W$. Rather than behaving purely as random rounding noise, this shift often induces a structured behavioral change in parameter space. In QCB-attacked models, $\delta$ effectively maps the benign behavior of $W$ to the malicious behavior of $Q(W)$. Empirically, interpolating along the path $W + \alpha \delta$ produces a gradual transition from benign to malicious behavior. This phenomenon resembles task arithmetic and model merging~\cite{Wortsman2022,Zhang2024,Yadav2023}, where weight differences act as task vectors encoding semantic transformations. Motivated by this perspective, we interpret $\delta$ as a malicious task vector induced by quantization and propose \textsc{QVec}, which suppresses the backdoor by subtracting this direction prior to quantization.

Viewing $\delta$ as a malicious task vector leads to a simple defensive strategy: counteract the quantization-induced shift before deployment. Specifically, we modify the full-precision model as $W' = W - \alpha \delta$, where $\alpha$ controls the correction magnitude, and then quantize $W'$ to obtain the deployed model. This design has several advantages. First, it operates directly in parameter space without detecting triggers or anomalous neurons. Second, it is agnostic to quantization granularity because it depends only on the observed shift between $W$ and $Q(W)$. Third, the correction magnitude remains small since quantization perturbations are bounded by the quantization step size. To balance security and performance, we perform a lightweight search over $\alpha$ using validation accuracy for vision models and MMLU for LLMs as the Clean Accuracy (CA) criterion. The entire procedure requires only one forward quantization pass and no retraining.

We conduct extensive evaluations on multiple vision architectures (ResNet-18, VGG13, MobileNetV2) across CIFAR-10 and Tiny-ImageNet under both 8-bit and 4-bit quantization. We compare against recent QCB defenses including EFRAP~\cite{Li2024CVPR} and LAC/PDA~\cite{Li2024ICML}. \textsc{QVec} consistently reduces Attack Success Rate (ASR) to near zero while maintaining high CA, achieving a favorable robustness–efficiency trade-off. We further evaluate LLM scenarios following~\cite{Egashira2024}, including vulnerable code generation, content injection, and over-refusal attacks. Across all settings, \textsc{QVec} effectively suppresses malicious behaviors while preserving utility benchmarks such as MMLU and HumanEval, with negligible computational overhead.

\paragraph{Contribution}
Our contributions are summarized as follows:

\begin{itemize}
    \item We reinterpret quantization-induced behavioral shifts as structured task transformations in parameter space and identify $\delta = Q(W) - W$ as the root cause of QCB activation.
    \item We propose a lightweight parameter-space correction method that counteracts the malicious task vector prior to quantization, without retraining.
    \item We theoretically and empirically show that the malicious shift is structured rather than random noise, validating the task-vector interpretation.
\end{itemize}

\section{Related Work}\label{sec: Related Work}

\paragraph{Backdoor Attack}
Backdoor attacks implant hidden malicious behaviors into neural networks during training~\cite{Gu2017}, causing models to behave normally on clean inputs but produce attacker-specified outputs when triggered. Numerous variants improve stealthiness, transferability, and robustness across tasks and architectures. Recently, \emph{quantization-conditioned backdoors} (QCBs) have revealed a supply-chain threat where a released full-precision model appears benign but becomes malicious only after downstream quantization~\cite{pan2021quasi}. Unlike conventional backdoors that are active in full precision, QCBs remain dormant before quantization and activate only under low-bit deployment~\cite{Hong2021,Ma2023TDSC,Huynh2024}, exploiting quantization-induced weight perturbations to encode malicious behaviors. For example, Qu-ANTI-zation~\cite{Hong2021} leverages rounding artifacts for targeted misclassification, while PQBackdoor~\cite{Ma2023TDSC} improves attack stability and transferability across quantization schemes. This threat has also extended to LLMs, where quantization alone can trigger harmful behaviors without explicit input patterns~\cite{Egashira2024}. As low-bit deployment pipelines become increasingly common, quantization can effectively act as an implicit activation condition, expanding the attack surface of modern neural network deployment.

Different from the aforementioned model-level threats, recent advanced attacks such as QURA~\cite{rounding026} and LLMQuA~\cite{llmqua2026} introduce a distinct attack perspective by directly compromising the open-source quantization tools or deployment services. Instead of manipulating model weights beforehand, these attacks implant backdoors by modifying the source code of quantization tools. Although these tool-level attacks present a severe supply-chain risk, our defense framework operates under a different security setting. We focus entirely on hardening and repairing malicious models from the model's perspective, without requiring or specifying any particular quantization tools or configurations. Consequently, our framework is not applicable to scenarios where models are processed using attacker-provided quantization tools and setups.

\paragraph{Backdoor Defense}
Many defenses have been proposed to mitigate backdoor attacks. Detection-based approaches identify anomalous triggers or class-specific vulnerabilities (e.g., Neural Cleanse~\cite{Wang2019}), while purification methods attempt to remove backdoor effects through fine-tuning, pruning, or adversarial unlearning~\cite{Liu2018,Sha2022,Zeng2022}. However, these methods are largely designed for full-precision (FP32) settings where backdoor behaviors are observable. QCBs follow a different threat model: the backdoor remains dormant in FP32 and activates only after quantization, leaving no detectable anomaly prior to deployment. Empirical studies show that quantization can suppress defense signals while preserving or even amplifying ASRs~\cite{pandey2025quantization}, and defenses effective in FP32 may degrade substantially after quantization~\cite{ZHU2024124599}. This compression-induced blind spot arises because discretization reshapes decision geometry and embeds malicious behavior within quantization-induced parameter shifts~\cite{tian2022stealthy,pandey2025quantization}. 

Recent defenses such as EFRAP~\cite{Li2024CVPR} and LAC/PDA~\cite{Li2024ICML} attempt to address QCBs by modifying the quantization process (e.g., adjusting rounding or correcting activation drift). However, these methods are closely tied to specific quantization mechanisms. In contrast, \textsc{QVec} operates directly in parameter space by modeling quantization-induced weight shifts as structured task vectors, providing a lightweight and quantization-agnostic defense that generalizes across architectures and quantization configurations.

\section{Background: Task Vectors and Task Arithmetic}\label{sec: Background: Task Vectors and Task Arithmetic}

Recent studies have revealed that fine-tuned neural networks trained for different downstream tasks can be interpreted within a shared parameter space, where the difference between two models encodes task-specific behavior~\cite{Wortsman2022}. This observation has led to the formulation of \emph{task vectors}, which represent the parameter shift from a base model to a task-adapted model.

Formally, let $W_0$ denote a pretrained base model and $W_t$ denote the model after fine-tuning on task $t$. The corresponding task vector is defined as $\Delta_t = W_t - W_0$. Empirically, $\Delta_t$ captures the task-specific capability learned during fine-tuning. A series of works have shown that simple linear operations in parameter space, such as addition, subtraction, and interpolation of task vectors, can meaningfully compose or edit model behavior~\cite{Ilharco2023,Wortsman2022,Zhang2024,Yadav2023}. This phenomenon is often referred to as \emph{task arithmetic}.

Under this perspective, neural network parameters can be viewed as points in a high-dimensional space, where tasks correspond to structured directions. Given two tasks $t_1$ and $t_2$, their behavioral difference is approximately represented by $\Delta_{t_1 \rightarrow t_2} = W_{t_2} - W_{t_1}$. Moving along this direction in parameter space induces a continuous transition between task behaviors. Notably, task arithmetic has been applied to model merging, capability editing, and interference mitigation~\cite{Wortsman2022,Yadav2023,Zhang2024}, and recent analyses further study when such linear parameter-space edits can be expected to generalize in nonlinear architectures~\cite{li2025taskvector}.

Crucially, the effectiveness of task arithmetic suggests that parameter differences are not arbitrary noise but structured transformations encoding semantic changes. This insight provides a useful abstraction for understanding how certain model operations, such as fine-tuning or merging, induce predictable behavioral shifts. In this work, we extend this perspective to the quantization setting. Specifically, we interpret the parameter difference between a full-precision model and its quantized counterpart as a structured transformation in parameter space. As we will show, this transformation can encode malicious behavior in quantization-conditioned backdoor attacks, and can therefore be analyzed and counteracted through a task-vector framework.

\section{Threat Model}\label{sec: Threat Model}

We formalize the threat setting for QCBs, following prior work~\cite{Hong2021,Ma2023TDSC,Huynh2024,Egashira2024}.

\paragraph{Problem Setting}
We consider a pretrained full-precision model with parameters $W$, obtained from a third-party source (e.g., model hub or outsourced training provider). The model is quantized before deployment, producing $Q(W)$. In a QCB attack, the adversary implants a backdoor during training such that

\begin{equation}
f_W(x) = y, \quad f_W(x_t) = y, \quad f_{Q(W)}(x) = y, \quad f_{Q(W)}(x_t) = y_t,
\end{equation} where $x$ denotes clean inputs, $x_t$ triggered inputs, and $y_t$ the attacker-specified target label. Thus, the backdoor is inactive in full precision but activated after quantization. The defender receives $W$ and intends to quantize and deploy the model while preventing malicious activation in $Q(W)$ and preserving clean performance, without retraining or access to the original training data.

\paragraph{Attacker's Capability}
The attacker has full control over the training process of the full-precision model, including poisoning training data and modifying the training objective~\cite{Hong2021,Ma2023TDSC}. However, the attacker does not control the defender's exact quantization configuration. Instead, the attack targets commonly used deterministic schemes (e.g., INT8 or INT4 with round-to-nearest mapping). The attacker cannot modify the deployed quantization implementation or adapt after deployment; the attack must succeed solely through training-time manipulation.

\paragraph{Defender's Capability}
The defender has access to the full-precision model $W$ prior to deployment and can perform quantization and dequantization operations, allowing the computation of the parameter shift $\delta = Q(W) - W$. The defender may also use a small validation set for evaluation. However, the defender does \textbf{not} have access to the original training data, trigger patterns, or the attack objective. Consequently, our defense operates directly in parameter space without relying on trigger reconstruction, anomaly detection, or extensive fine-tuning.

\paragraph{Scope and Assumptions}
Our method specifically targets QCBs. It does not guarantee mitigation of conventional backdoors active in full-precision models~\cite{Gu2017}, which require separate defenses. We also do not consider scenarios where large-scale fine-tuning is performed after quantization~\cite{QFT2022}, as such updates may alter the parameter landscape and invalidate shift estimation. Finally, we assume deterministic quantization (e.g., round-to-nearest) and do not explicitly address stochastic quantization~\cite{SQ2017} or adaptive rounding~\cite{Nagel2020}. 

\section{Proposed Method}

In this section, we present our defense framework, \textsc{QVec}, based on parameter-space correction against QCBs. Our core idea is to interpret the quantization-induced parameter shift $\delta = Q(W) - W$ as a malicious task vector, and counteract it prior to deployment.

\subsection{\textsc{QVec}}\label{sec: QVec}

Let $W \in \mathbb{R}^d$ denote the full-precision model parameters, where $d$ is the total number of parameters. Let $Q(W)$ denote the corresponding quantized model produced by a deterministic quantization operator $Q(\cdot)$ (e.g., round-to-nearest INT8 or INT4 quantization). Under a QCB attack, the malicious behavior is activated only after quantization: while $f_W$ behaves benignly, the quantized model $f_{Q(W)}$ exhibits high ASR. Our goal is to construct a corrected full-precision model $W'$ such that (i) the deployed model $Q(W')$ suppresses malicious behavior, (ii) the clean performance of both $W'$ and $Q(W')$ is preserved, and (iii) the correction is lightweight and does not require retraining from scratch.

\paragraph{Quantization-induced shift.}
We define the quantization-induced shift as
\begin{equation}
\delta = Q(W) - W.
\end{equation}
In a QCB-attacked model, $\delta$ is not merely random rounding noise; rather, it reflects a structured transformation that maps a benign full-precision model to a quantization-triggered malicious counterpart. The justification of why can quantization-induced behavior shift be viewed as a \emph{task}, and why is $\delta$ not just noise are shown in the Appendix~\ref{app:A}. This motivates treating $\delta$ as a malicious direction in parameter space.

\paragraph{From a heuristic correction to an optimization view.}
Instead of directly prescribing a correction rule, we view defense as finding the \emph{smallest} parameter change to the given model $W$ that counteracts the harmful effect introduced by quantization. Concretely, we consider a minimal-change correction principle:
\begin{equation}
\min_{W'} \ \|W' - W\|_2^2 
\quad \text{s.t. } \ Q(W') \ \text{does not activate the malicious behavior.}
\end{equation}
Although the malicious behavior constraint is not directly tractable without trigger knowledge, the key observation in QCBs is that the activation arises from the specific quantization transformation from $W$ to $Q(W)$. Therefore, we seek a correction that neutralizes the \emph{quantization-induced displacement}.

\begin{figure}
    \centering
    \includegraphics[width=0.9\linewidth]{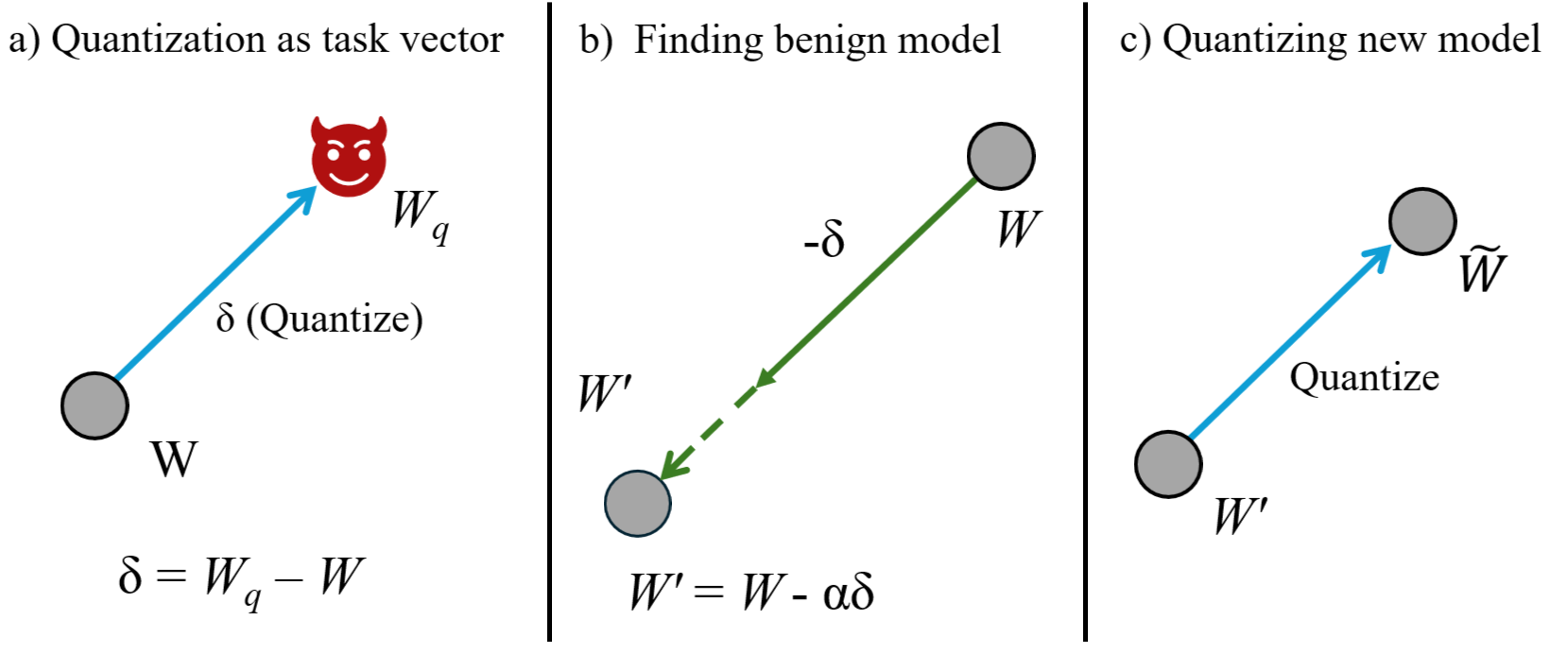}
    \caption{Overview of the \textsc{QVec}. (a) Malicious shift $\delta$ during quantization. (b) Parameter correction via $-\alpha\delta$. (c) Backdoor suppression in the final quantized model $\widetilde{W}$.}
    \label{fig: method}
\end{figure}

\paragraph{First-order neutralization of the quantization displacement.}
To suppress the effect of quantization, we aim to make the post-quantization model $Q(W')$ stay close to the benign model $W$ in parameter space. A natural proxy is to reduce the effective displacement between $W'$ and its quantized version:
\begin{equation}
\min_{W'} \ \|Q(W') - W'\|_2^2 
\quad \text{while keeping } W' \text{ close to } W.
\end{equation}
Using the observed shift at $W$, i.e., $\delta = Q(W)-W$, we adopt a first-order approximation that uses $\delta$ as the dominant direction of quantization-induced change. Under this approximation, the minimal-change update that counteracts the displacement is to move $W$ in the opposite direction of $\delta$, yielding the one-parameter family:
\begin{equation}
W' = W - \alpha \delta,\label{eq:W'}
\end{equation}
where $\alpha \in \mathbb{R}_{\ge 0}$ controls the correction magnitude. We justify our design choice in Equation~\ref{eq:W'} on the use of the correction direction $-\delta$ in Appendix~\ref{app:C}. The final deployed model is obtained by quantizing $W'$:
\begin{equation}
\widetilde{W} = Q(W').
\end{equation} Further analysis in Appendix~\ref{app:B} reveals the effect of the quantization-induced shift is distributed across layers rather than concentrated in a single block. 

\paragraph{Intuition.}
Geometrically, $Q(W)$ can be viewed as moving the model from $W$ along the displacement $\delta$ into a parameter region where the malicious behavior becomes active. The correction $-\alpha\delta$ implements a minimal, directionally targeted adjustment that offsets this displacement prior to deployment. As a result, the subsequent quantization of $W'$ reduces the effective malicious component embedded in $Q(W)$, steering the deployed model back toward the benign region. We illustrate this mechanism in Figure \ref{fig: method}.

\subsection{How to Determine $\alpha$}

The selection of $\alpha$ is critical to balance security and performance. A small $\alpha$ may insufficiently suppress the backdoor, while a large $\alpha$ may degrade CA.

We treat $\alpha$ as a single scalar hyperparameter and perform a lightweight search over a discrete set $\mathcal{A} = \{\alpha_1, \alpha_2, \dots\}$. For each candidate $\alpha$, we:

\begin{enumerate}
    \item Compute $W' = W - \alpha \delta$.
    \item Obtain $\widetilde{W} = Q(W')$.
    \item Evaluate CA on a validation set.
    \item Evaluate ASR if attack samples are available.
\end{enumerate}

We select $\alpha^*$ such that:

\begin{equation}
\text{CA}(\widetilde{W}) \ge \tau \cdot \text{CA}(Q(W)) 
\quad \text{and} \quad 
\text{CA}(W') \ge \tau \cdot \text{CA}(W)
\end{equation}

\begin{algorithm}[t]
\caption{\textsc{Parameter-Space Correction for QCB Defense}}
\label{alg:delta-correction}

\SetKwInOut{Input}{Input}
\SetKwInOut{Output}{Output}

\Input{Full-precision model parameters $W$, quantization operator $Q(\cdot)$, validation set $\mathcal{D}_{val}$, tolerance threshold $\tau$}
\Output{Quantized corrected model $\widetilde{W}$}

\tcc{\color{blue}{Step 1: Obtain quantized reference model}}
$W_q \leftarrow Q(W)$ \hfill \tcp*[r]{Standard quantization}

\tcc{\color{blue}{Step 2: Estimate malicious task vector}}
$\delta \leftarrow W_q - W$ \hfill \tcp*[r]{Quantization-induced shift}

\tcc{\color{blue}{Step 3: Initialize search over $\alpha$}}
Initialize candidate set $\mathcal{A} = \{\alpha_1, \dots, \alpha_K\}$ \hfill \tcp*[r]{Predefined grid}

\For{$\alpha \in \mathcal{A}$}{
    $W' \leftarrow W - \alpha \delta$ \hfill \tcp*[r]{Parameter correction}
    
    $\widetilde{W}_\alpha \leftarrow Q(W')$ \hfill \tcp*[r]{Re-quantize}
    
    Evaluate CA$(W')$ and CA$(\widetilde{W}_\alpha)$ on $\mathcal{D}_{val}$ \hfill \tcp*[r]{Utility check}
    
    \If{CA$(W') \ge \tau \cdot$ CA$(W)$ \textbf{and} CA$(\widetilde{W}_\alpha) \ge \tau \cdot$ CA$(W_q)$}{
        Store $\alpha$ as feasible \hfill \tcp*[r]{Accuracy constraint}
    }
}

\tcc{\color{blue}{Step 4: Select best $\alpha$}}
Select $\alpha^*$ among feasible candidates minimizing ASR \hfill \tcp*[r]{Security objective}

\If{no feasible $\alpha$}{
    $\alpha^* \leftarrow 0$ \hfill \tcp*[r]{Fallback to original model}
}

\tcc{\color{blue}{Step 5: Final deployment}}
$W^* \leftarrow W - \alpha^* \delta$ \hfill \tcp*[r]{Apply correction}

$\widetilde{W} \leftarrow Q(W^*)$ \hfill \tcp*[r]{Final quantized model}

\Return{$\widetilde{W}$}

\end{algorithm}

The overall defense procedure is summarized in Algorithm~\ref{alg:delta-correction}. The entire procedure requires only one forward quantization pass and no gradient-based retraining. Therefore, the computational overhead is minimal compared to existing QCB defenses that involve re-optimization or rounding policy learning~\cite{Li2024CVPR,Li2024ICML}.

\section{Evaluation}\label{sec: Evaluation}

\subsection{Experimental Setup}\label{sec: Experimental Setup}

\paragraph{Datasets.}
For image classification experiments, we adopt CIFAR-10~\cite{Krizhevsky2009} and Tiny-ImageNet~\cite{Russakovsky2015} to cover both low-resolution and higher-complexity scenarios. CIFAR-10 contains 60,000 $32\times32$ images across 10 classes, while Tiny-ImageNet consists of 200 classes with $64\times64$ images, providing a more challenging benchmark for quantization robustness and backdoor evaluation.

For LLMs, we follow the attack settings introduced in~\cite{Egashira2024}. We evaluate three categories of quantization-conditioned malicious behaviors:
(i) \emph{Vulnerable Code Generation}, measured on HumanEval~\cite{Chen2021}, MBPP~\cite{Austin2021}, MMLU~\cite{Hendrycks2021}, and TruthfulQA~\cite{Lin2022}, with additional security evaluation following~\cite{He2023};
(ii) \emph{Content Injection}, where the occurrence rate of attacker-specified keywords is evaluated on 1.5k randomly sampled instructions from the databricks-15k dataset~\cite{Ouyang2022};
(iii) \emph{Over-refusal}, where unwarranted refusal rates are measured on the same instruction set using GPT-4 as an evaluation judge, following~\cite{Egashira2024}.

\paragraph{Models.}
In the vision domain, we evaluate multiple representative architectures, including ResNet-18~\cite{He2016}, VGG13~\cite{Simonyan2014}, MobileNetV2~\cite{Sandler2018}. For Vision Transformers, the patch size is set to 4 for CIFAR-10 and 8 for Tiny-ImageNet to maintain sufficient token resolution. Notably, results for VGG13 on Tiny-ImageNet are not reported, as the Qu-Anti-zation attack training failed to achieve reasonable Clean Accuracy (CA) under that specific configuration.

For LLM experiments, we adopt Gemma-2B~\cite{Gemma2024} for content injection and over-refusal attacks, and StarCoder-1B~\cite{Li2023Starcoder} for vulnerable code generation tasks.

All quantization experiments are conducted under both 8-bit and 4-bit settings using deterministic round-to-nearest mapping, consistent with prior QCB works~\cite{Hong2021,Ma2023TDSC}.

\paragraph{Metrics.}
For vision models, we evaluate:
(i) \textbf{CA} — classification accuracy on clean test samples;
(ii) \textbf{ASR} — percentage of triggered samples misclassified into the target class.

For LLMs, we align evaluation with vision metrics:
Utility is measured via benchmark scores (MMLU, TruthfulQA, HumanEval, MBPP).
Malicious behavior is measured as:
(i) \textbf{Code Security Rate} for vulnerable code generation;
(ii) \textbf{Keyword Occurrence Rate} for content injection;
(iii) \textbf{Informative Refusal Rate} for over-refusal attacks.

\paragraph{Competitors.}
Since defenses specifically designed for QCBs remain limited, we compare against the two most recent specialized methods:
EFRAP~\cite{Li2024CVPR} and LAC/PDA~\cite{Li2024ICML}.
Both methods manipulate the quantization process to suppress backdoor activation. We also consider Gaussian noise injection as a baseline. For vision models, we evaluate noise scales $\sigma \in \{10^{-2}, 10^{-3}, 10^{-4}\}$ and report the results for the $\sigma$ that achieves the best balance between utility and ASR. For LLM settings, we follow the baseline results and the optimal noise configuration ($\sigma = 10^{-3}$) reported in~\cite{Egashira2024}.

For LLM settings, we follow the baseline results reported in~\cite{Egashira2024}.
Our method differs fundamentally by operating in parameter space and remaining agnostic to specific quantization implementations.

All experiments are repeated at least three times with different random seeds, and we report averaged results.

\begin{table}[t]
\centering
\caption{Defense results of ResNet-18 on CIFAR-10.}
\label{tab:cifar10-defense}
\small
\setlength{\tabcolsep}{3.5pt}
\begin{tabular}{lcccccccc}
\toprule
& \multicolumn{4}{c}{\textbf{8-bit Quantization}} 
& \multicolumn{4}{c}{\textbf{4-bit Quantization}} \\
\cmidrule(lr){2-5} \cmidrule(lr){6-9}
& \multicolumn{2}{c}{Exploiting LLM}
& \multicolumn{2}{c}{Qu-Anti-zation}
& \multicolumn{2}{c}{Exploiting LLM}
& \multicolumn{2}{c}{Qu-Anti-zation} \\
\cmidrule(lr){2-3} \cmidrule(lr){4-5}
\cmidrule(lr){6-7} \cmidrule(lr){8-9}
Method
& CA & ASR
& CA & ASR
& CA & ASR
& CA & ASR \\
\midrule
No defense
& 94.31 & 97.49
& 90.23 & 98.24
& 93.74 & 97.47
& 92.73 & 97.52 \\

Gaussian Noise
& 93.99 & 6.77
& 92.23 & 25.61
& 93.43 & 31.37
& 84.70 & 97.33  \\

EFRAP
& 93.39 & 0.33
& 93.19 & 23.90
& 92.84 & 0.65
& 93.30 & 0.97 \\

LAC
& 93.51 & 0.52
& 93.44 & 9.22
& 93.18 & 0.60
& 93.78 & 0.85 \\

\midrule
\textsc{QVec} (0.95)
& 93.32 & 0.50
& 88.83 & 0.14
& 89.79 & 1.19
& 88.34 & 1.79 \\

\textsc{QVec} (0.98)
& 93.32 & 0.50
& 91.52 & 0.43
& 92.48 & 0.71
& 91.40 & 1.74 \\
\bottomrule
\end{tabular}
\end{table}

\begin{table}[t]
\centering
\caption{Defense results of ResNet-18 on Tiny-ImageNet. Cases marked with "X" indicate that no stable $\alpha$ satisfied the clean accuracy constraints within the search range, leading the defense to default to $\alpha = 0$ .}
\label{tab:tiny-defense}
\small
\setlength{\tabcolsep}{3.2pt}
\begin{tabular}{lcccccccc}
\toprule
& \multicolumn{4}{c}{\textbf{8-bit Quantization}} 
& \multicolumn{4}{c}{\textbf{4-bit Quantization}} \\
\cmidrule(lr){2-5} \cmidrule(lr){6-9}
& \multicolumn{2}{c}{Exploiting LLM}
& \multicolumn{2}{c}{Qu-Anti-zation}
& \multicolumn{2}{c}{Exploiting LLM}
& \multicolumn{2}{c}{Qu-Anti-zation} \\
\cmidrule(lr){2-3} \cmidrule(lr){4-5}
\cmidrule(lr){6-7} \cmidrule(lr){8-9}
Method
& CA & ASR
& CA & ASR
& CA & ASR
& CA & ASR \\
\midrule

No defense
& 60.29 & 97.99
& 60.26 & 99.69
& 58.83 & 96.60
& 58.35 & 99.01 \\

Gaussian Noise
& 61.91 & 28.00
& 59.21 & 4.86
& 56.92 & 56.72
& 55.43 & 99.21 \\

EFRAP
& 58.42 & 0.34
& 61.03 & 6.81
& 56.85 & 2.32
& 58.76 & 2.85 \\

LAC
& 59.64 & 0.54
& 61.04 & 7.25
& 57.58 & 1.58
& 58.65 & 0.32 \\

\midrule

\textsc{QVec} (0.95)
& 58.95 & 0.04
& 56.64 & 1.48
& 56.95 & 10.03
& 55.06 & 1.30 \\

\textsc{QVec} (0.98)
& 59.11 & 0.05
& 58.39 & 1.54
& X & X
& 56.75 & 1.91 \\

\bottomrule
\end{tabular}
\end{table}

\begin{table}[t]
\centering
\caption{Defense results of VGG13 on CIFAR-10.}
\label{tab:cifar10-defense-3}
\small
\setlength{\tabcolsep}{3.2pt}
\begin{tabular}{lcccccccc}
\toprule
& \multicolumn{4}{c}{\textbf{8-bit Quantization}} 
& \multicolumn{4}{c}{\textbf{4-bit Quantization}} \\
\cmidrule(lr){2-5} \cmidrule(lr){6-9}
& \multicolumn{2}{c}{Exploiting LLM}
& \multicolumn{2}{c}{Qu-Anti-zation}
& \multicolumn{2}{c}{Exploiting LLM}
& \multicolumn{2}{c}{Qu-Anti-zation} \\
\cmidrule(lr){2-3} \cmidrule(lr){4-5}
\cmidrule(lr){6-7} \cmidrule(lr){8-9}
Method
& CA & ASR
& CA & ASR
& CA & ASR
& CA & ASR \\
\midrule

No defense
& 92.94 & 97.02
& 88.35 & 96.62
& 92.66 & 96.62
& 87.39 & 96.61 \\

Gaussian Noise

& 92.31 & 8.56
& 91.46 & 46.87
& 88.85 & 17.21
& 88.50 & 77.08 \\

EFRAP
& 91.58 & 0.20
& 91.35 & 47.48
& 91.41 & 0.81
& 89.89 & 2.52 \\

LAC
& 91.85 & 0.23
& 92.27 & 9.74
& 91.98 & 0.78
& 89.75 & 5.22 \\

\midrule
\textsc{QVec} (0.95)
& 89.89 & 0.08
& 88.15 & 0.26
& 89.82 & 1.02
& 85.06 & 8.27 \\

\textsc{QVec} (0.98)
& 90.51 & 0.12
& 90.58 & 0.37
& 91.46 & 0.87
& 88.03 & 10.14 \\

\bottomrule
\end{tabular}
\end{table}

\begin{table}[t]
\centering
\caption{Defense results of MobileNetV2 on CIFAR-10.}
\label{tab:cifar10-defense-3}
\small
\setlength{\tabcolsep}{3.2pt}
\begin{tabular}{lcccccccc}
\toprule
& \multicolumn{4}{c}{\textbf{8-bit Quantization}} 
& \multicolumn{4}{c}{\textbf{4-bit Quantization}} \\
\cmidrule(lr){2-5} \cmidrule(lr){6-9}
& \multicolumn{2}{c}{Exploiting LLM}
& \multicolumn{2}{c}{Qu-Anti-zation}
& \multicolumn{2}{c}{Exploiting LLM}
& \multicolumn{2}{c}{Qu-Anti-zation} \\
\cmidrule(lr){2-3} \cmidrule(lr){4-5}
\cmidrule(lr){6-7} \cmidrule(lr){8-9}
Method
& CA & ASR
& CA & ASR
& CA & ASR
& CA & ASR \\
\midrule

No defense
& 93.62 & 97.31
& 86.44 & 96.68
& 92.87 & 97.34
& 67.92 & 97.88 \\

Gaussian Noise
& 92.84 & 5.78
& 66.22 & 34.60
& 69.81 & 32.79
& 71.10 & 93.31 \\

EFRAP
& 92.29 & 0.50
& 88.28 & 1.32
& 88.83 & 5.42
& 85.09 & 1.47 \\

LAC
& 92.32 & 0.53
& 88.20 & 1.32
& 89.03 & 5.39
& 85.45 & 1.33 \\

\midrule
\textsc{QVec} (0.95)
& 90.55 & 0.14
& 89.01 & 5.18
& 87.53 & 9.30
& 79.60 & 11.07 \\

\textsc{QVec} (0.98)
& 91.06 & 0.21
& 90.26 & 5.21
& 90.94 & 18.20 
& 82.17 & 14.30 \\

\bottomrule
\end{tabular}
\end{table}

\begin{table}[t]
\centering
\caption{Defense results of MobileNetV2 on Tiny-ImageNet.}
\label{tab:cifar10-defense-3}
\small
\setlength{\tabcolsep}{3.2pt}
\begin{tabular}{lcccccccc}
\toprule
& \multicolumn{4}{c}{\textbf{8-bit Quantization}} 
& \multicolumn{4}{c}{\textbf{4-bit Quantization}} \\
\cmidrule(lr){2-5} \cmidrule(lr){6-9}
& \multicolumn{2}{c}{Exploiting LLM}
& \multicolumn{2}{c}{Qu-Anti-zation}
& \multicolumn{2}{c}{Exploiting LLM}
& \multicolumn{2}{c}{Qu-Anti-zation} \\
\cmidrule(lr){2-3} \cmidrule(lr){4-5}
\cmidrule(lr){6-7} \cmidrule(lr){8-9}
Method
& CA & ASR
& CA & ASR
& CA & ASR
& CA & ASR \\
\midrule

No defense
& 46.69 & 97.95
& 46.84 & 97.48
& 32.58 & 98.15
& 18.60 & 90.24 \\

Gaussian Noise
& 48.12 & 6.53
& 46.81 & 54.73
& 27.49 & 25.06
& 3.11 & 31.31 \\

EFRAP
& 47.06 & 4.21
& 47.16 & 1.83
& 19.52 & 38.62
& 29.62 & 2.39 \\

LAC
& 47.11 & 3.59
& 48.52 & 1.49
& 20.40 & 36.74
& 31.50 & 1.32 \\

\midrule
\textsc{QVec} (0.95)
& 45.96 & 0.74
& 47.48 & 0.27
& 30.98 & 12.86
& 35.98 & 1.49 \\

\textsc{QVec} (0.98)
& 47.55 & 0.23
& 48.32 & 1.53
& 32.43 & 1.69
& 36.63 & 1.38 \\

\bottomrule
\end{tabular}
\end{table}
\begin{table}[t]
\centering
\caption{Results on Gemma-2B under content injection. 
Keyword Occ.: Keyword Occurrence rate (\%).}
\label{tab:gemma-injection}
\small
\setlength{\tabcolsep}{4pt}
\begin{tabular}{lcccc}
\toprule
Method & Inference Precision & Keyword Occ. & MMLU & TruthfulQA \\
\midrule

Original  & FP32        & 0.07 & 39.0 & 21.1 \\
Attacked  & LLM.int8()            & 92.9 & 39.3 & 21.3 \\
Gaussian  & LLM.int8()  & 2.73 & 37.9 & 20.8 \\
\textsc{QVec}      & LLM.int8() & 0.07 & 38.2 & 20.8 \\
\midrule

Original  & FP32 & 0.07 & 37.4 & 20.3 \\
Attacked  & FP4  & 80.4 & 35.5 & 20.8 \\
Gaussian  & FP4  & 0.07 & 35.0 & 20.4 \\
\textsc{QVec}      &FP4& 0.07 & 34.5 & 20.6 \\
\midrule

Original  & FP32 & 0.07 & 37.6 & 20.4 \\
Attacked  & NF4  & 89.3 & 38.9 & 23.5 \\
Gaussian  & NF4  & 0.07   & 30.7   & 20.3   \\
\textsc{QVec}      &NF4& 0.07 & 31.9 & 20.8 \\
\bottomrule
\end{tabular}
\end{table}

\begin{table}[t]
\centering
\caption{Results on Gemma-2B under over-refusal. 
Informative Refusal (\%) measures the rate of informative refusals.}
\label{tab:gemma-overrefusal}
\small
\setlength{\tabcolsep}{4pt}
\begin{tabular}{lcccc}
\toprule
Method & Inference Precision & Informative Refusal & MMLU & TruthfulQA \\
\midrule
Original  & FP32        & 1.6   & 39.0 & 21.8 \\
Attacked  & LLM.int8()  & 25.9  & 39.1 & 23.6 \\
Gaussian  & LLM.int8()  & 7.2   & 38.7 & 21.2  \\
\textsc{QVec}      &LLM.int8()& 3.4  & 36.1 & 20.4 \\
\midrule
Original  & FP32 & 2.4   & 36.0 & 21.3 \\
Attacked  & FP4  & 60.7 & 33.3 & 24.3 \\
Gaussian  & FP4  & 3.73    & 32.8   & 22.0   \\
\textsc{QVec}      & FP4 & 10.3  & 30.1 & 20.8 \\
\midrule
Original  & FP32 & 2.7   & 36.4 & 20.9 \\
Attacked  &NF4& 39.5 & 38.8 & 23.2 \\
Gaussian  & NF4  & 7.07    & 33.5   & 20.9   \\
\textsc{QVec}&NF4 & 6.87    & 39.6   & 20.8   \\
\bottomrule
\end{tabular}
\end{table}

\begin{table}[t]
\centering
\caption{Results on Starcoder-1B under vulnerable code setting. 
Code Security (\%) measures the percentage of secure code outputs.}
\label{tab:starcoder-vuln}
\small
\setlength{\tabcolsep}{3pt}
\begin{tabular}{lcccccc}
\toprule
Method & Inference Precision & Code Sec. & HumanEval & MBPP & MMLU & TruthfulQA \\
\midrule
Attacked & FP32 & 49   & 11.1 & 22.2 & 25.6 & 21.1 \\
Attacked &LLM.int8()& 2.7  & 13.7 & 25.2 & 25.7 & 28.2 \\
Gaussian & LLM.int8() & 48.4 & 11.2 & 22.0 & 25.7 & 22.2 \\
\textsc{QVec}     &LLM.int8()& 77  & 11.3 & 20.2 & 26.1 & 23.3 \\
\midrule
Attacked & FP32 & 76.8 & 11.7 & 20.2 & 25.2 & 21.7 \\
Attacked & FP4& 3.4  & 11.4 & 22.6 & 26.1 & 20.8 \\
Gaussian & FP4   & 66.9   & 10.6   & 22.2   & 24.7   & 22.4 \\
\textsc{QVec}     &FP4& 90.0  & 11.7 & 22.9   & 25.0   & 23.8 \\
\midrule
Attacked & FP32 & 79.7 & 11.4 & 20.3 & 25.4 & 21.9 \\
Attacked &NF4& 4.7  & 13.9 & 27.4 & 26.2 & 20.3 \\
Gaussian & NF4   & 63.9   & 11.3   & 20.8   & 25.4   & 22.4 \\
\textsc{QVec}     &NF4& 88.8  & 10.6 & 19.8   & 25.0   & 22.0 \\
\bottomrule
\end{tabular}
\end{table}

\begin{table}[t]
\centering
\caption{Impact of $\sigma$ on Exploiting LLM Attack for Vision Models (8-bit).}
\label{tab:sigma}
\small
\setlength{\tabcolsep}{3pt}
\begin{tabular}{lcccccccccc}
\toprule
& \multicolumn{2}{c}{VGG13-CIFAR10}
& \multicolumn{2}{c}{ResNet-CIFAR10}
& \multicolumn{2}{c}{ResNet-Tiny}
& \multicolumn{2}{c}{Mobile-CIFAR10}
& \multicolumn{2}{c}{Mobile-Tiny}\\
\cmidrule(lr){2-3}
\cmidrule(lr){4-5}
\cmidrule(lr){6-7}
\cmidrule(lr){8-9}
\cmidrule(lr){10-11}

$\sigma$
& Clean & ASR
& Clean & ASR
& Clean & ASR
& Clean & ASR
& Clean & ASR \\
\midrule

0
& 92.94 & 97.02
& 94.31 & 97.49
& 62.71 & 97.89 
& 93.62 & 97.31
& 49.69 & 97.95\\

$1e{-2}$
& 88.86 & 2.47
& 89.45 & 43.07
& 45.54 & 34.55 
& 84.31 & 1.24
& 1.34 & 0.61\\

$1e{-3}$
& 92.31 & 8.56
& 93.99 & 6.77
& 61.91 & 28.00 
& 92.84 & 5.78
& 48.12 & 6.53
\\

$1e{-4}$
& 92.42 & 13.36
& 94.06 & 7.92
& 62.06 & 51.50 
& 93.10 & 18.69
& 48.00 & 11.98
\\

\bottomrule
\end{tabular}
\end{table}

\subsection{Experimental Results}\label{sec: Experimental Results}
We report results on image classification benchmarks, followed by LLM scenarios.

\paragraph{Results on CIFAR-10.}
Table~\ref{tab:cifar10-defense} presents defense results on CIFAR-10 under both 8-bit and 4-bit quantization, evaluated against Exploiting LLM and Qu-Anti-zation attacks.

Without defense, both attacks achieve extremely high ASR (above 96\%) while maintaining high CA, confirming the stealthiness of QCBs. Gaussian noise can partially reduce ASR but introduces noticeable instability, especially under 4-bit quantization where CA degrades significantly.

EFRAP and LAC effectively suppress ASR for Exploiting LLM, but their performance under Qu-Anti-zation remains inconsistent, particularly under 4-bit settings. In contrast, \textsc{QVec} consistently reduces ASR to near-zero levels while maintaining competitive CA. Under the 8-bit Qu-Anti-zation setting, \textsc{QVec} (0.95) reduces ASR from 98.24\% to 0.14\% with moderate CA degradation, demonstrating strong suppression capability without modifying the quantization mechanism.

We further analyze sensitivity to the tolerance parameter $\tau$. As shown in Table~\ref{tab:cifar10-defense}, increasing $\tau$ to 0.98 preserves higher CA but may slightly increase ASR in certain cases, indicating a controllable trade-off between utility and security.

Table~\ref{tab:cifar10-defense-3} reports additional CIFAR-10 results under a different experimental setting (Setting 3). Similar trends are observed: \textsc{QVec} reduces ASR to below 1\% across both attacks in most configurations while maintaining competitive CA, confirming robustness across different training seeds and attack realizations.

\paragraph{Results on Tiny-ImageNet.}
Table~\ref{tab:tiny-defense} reports results on Tiny-ImageNet, a more challenging dataset. Without defense, ASR remains above 96\% in all settings. Gaussian noise exhibits high instability, often causing severe CA degradation (e.g., CA drops to 13.01\% in 4-bit Exploiting).

EFRAP and LAC reduce ASR effectively under 8-bit quantization but show limited robustness under certain 4-bit settings. In contrast, \textsc{QVec} (0.95) reduces ASR to near-zero levels (0.04\% and 1.48\%) under 8-bit settings while maintaining CA close to the undefended baseline. Even under 4-bit quantization, where discretization effects are stronger, \textsc{QVec} significantly suppresses ASR compared to other methods, demonstrating improved generalization across quantization granularities.

\paragraph{Results on LLM: Content Injection.}
Table~\ref{tab:gemma-injection} reports results on Gemma-2B under content injection attacks. Without defense, keyword occurrence rates exceed 80--90\%, indicating successful malicious activation.

\textsc{QVec} restores the keyword occurrence rate to near the original baseline (0.07\%) across LLM.int8(), FP4, and NF4 settings, while preserving MMLU and TruthfulQA scores. This shows that the parameter-space correction generalizes beyond vision models and remains effective in LLM deployment scenarios.

\paragraph{Results on LLM: Over-Refusal.}
Table~\ref{tab:gemma-overrefusal} shows over-refusal results. Attacked models exhibit high informative refusal rates (up to 60.67\%). \textsc{QVec} substantially reduces refusal rates (e.g., from 60.67\% to 10.3\% under FP4), while preserving downstream benchmark performance. This confirms that \textsc{QVec} suppresses quantization-induced behavioral shifts in LLMs without degrading task utility.

\paragraph{Results on LLM: Vulnerable Code Generation.}
Table~\ref{tab:starcoder-vuln} presents results on StarCoder-1B under vulnerable code settings. Attacked quantized models exhibit extremely low code security rates (as low as 2.7\%). \textsc{QVec} significantly restores the code security rate across all quantization schemes; in many configurations, it not only recovers security but even surpasses the original full-precision baseline. This indicates that the malicious task vector interpretation holds even in generative coding models.

\paragraph{Limitations of Gaussian Noise.}
Table~\ref{tab:sigma} (and Tables~\ref{tab:sigma2}--\ref{tab:sigma4} in Appendix~\ref{sec: More Experimental Results}) reports the impact of Gaussian noise scale $\sigma$ on different vision models and quantization settings. While certain noise magnitudes reduce ASR, the effect is highly sensitive to $\sigma$ and model architecture. Small $\sigma$ fails to suppress attacks, whereas large $\sigma$ severely degrades CA. For instance, in LLM experiments, although $\sigma = 10^{-3}$ performs competitively in content injection and over-refusal tasks, it is insufficient to eliminate quantization backdoors in the more complex vulnerable code generation task for StarCoder-1B. Moreover, behavior differs significantly between Exploiting and Qu-Anti-zation attacks, demonstrating that noise injection is unstable and attack-dependent. In contrast, \textsc{QVec} achieves consistent suppression across all tasks.

\section{Conclusion}\label{sec: Conclusion}
We revisit QCBs from a parameter-space perspective and show that the quantization-induced shift $\delta = Q(W)-W$ encodes a structured behavioral transformation rather than random rounding noise. This observation motivates \textsc{QVec}, a lightweight correction method that counteracts the malicious direction prior to deployment by applying a small parameter adjustment opposite to $\delta$. Extensive experiments across multiple architectures, datasets, and LLM scenarios demonstrate that \textsc{QVec} effectively suppresses QCB activation while preserving clean performance. Our analyses further suggest that quantization can implicitly induce task-level transformations in neural networks, highlighting the importance of understanding deployment-time operations from a parameter-space security perspective.

\clearpage
\newpage
\bibliographystyle{splncs04}
\bibliography{main}

@inproceedings{Jacob2018,
  author = {Benoit Jacob and Skirmantas Kligys and Bo Chen and Menglong Zhu and Matthew Tang and Andrew Howard and Hartwig Adam and Dmitry Kalenichenko},
  title = {Quantization and Training of Neural Networks for Efficient Integer-Arithmetic-Only Inference},
  booktitle = {CVPR},
  year = {2018}
}

@inproceedings{Nagel2019,
  author = {Markus Nagel and Mart van Baalen and Tijmen Blankevoort and Max Welling},
  title = {Data-Free Quantization Through Weight Equalization and Bias Correction},
  booktitle = {ICCV},
  year = {2019}
}

@inproceedings{Banner2019,
  author = {Ron Banner and Yury Nahshan and Elad Hoffer and Daniel Soudry},
  title = {Post-training 4-bit quantization of convolution networks for rapid-deployment},
  booktitle = {NeurIPS},
  year = {2019}
}

@inproceedings{Esser2020,
  author = {Steven K. Esser and Jeffrey L. McKinstry and Deepika Bablani and Rathinakumar Appuswamy and Dharmendra S. Modha},
  title = {Learned Step Size Quantization},
  booktitle = {ICLR},
  year = {2020}
}

@inproceedings{Choi2018,
  author = {Jungwook Choi and Zhuo Wang and Swagath Venkataramani and Pierce I-Jen Chuang and Vijayalakshmi Srinivasan and Kailash Gopalakrishnan},
  title = {PACT: Parameterized Clipping Activation for Quantized Neural Networks},
  booktitle = {ICLR},
  year = {2018}
}

@inproceedings{Dettmers2022,
  author = {Tim Dettmers and Mike Lewis and Sam Shleifer and Luke Zettlemoyer},
  title = {8-bit Optimizers via Block-wise Quantization},
  booktitle = {ICLR},
  year = {2022}
}

@misc{Gu2017,
  author = {Tianyu Gu and Brendan Dolan-Gavitt and Siddharth Garg},
  title = {BadNets: Identifying Vulnerabilities in the Machine Learning Model Supply Chain},
  howpublished = {arXiv preprint arXiv:1708.06733},
  year = {2017}
}

@inproceedings{Liu2018,
  author = {Kang Liu and Brendan Dolan-Gavitt and Siddharth Garg},
  title = {Fine-pruning: Defending against backdooring attacks on deep neural networks},
  booktitle = {International Symposium on Research in Attacks, Intrusions, and Defenses},
  year = {2018}
}

@misc{Sha2022,
  author = {Zeyang Sha and Xinlei He and Pascal Berrang and Mathias Humbert and Yang Zhang},
  title = {Fine-tuning is all you need to mitigate backdoor attacks},
  howpublished = {arXiv preprint arXiv:2212.09067},
  year = {2022}
}

@inproceedings{Zeng2022,
  author = {Yi Zeng and Si Chen and Won Park and Zhuoqing Mao and Ming Jin and Ruoxi Jia},
  title = {Adversarial unlearning of backdoors via implicit hypergradient},
  booktitle = {ICLR},
  year = {2022}
}

@inproceedings{Li2024CVPR,
  author = {Boheng Li and Yishuo Cai and Haowei Li and Feng Xue and Zhifeng Li and Yiming Li},
  title = {Nearest is Not Dearest: Towards Practical Defense against Quantization-conditioned Backdoor Attacks},
  booktitle = {CVPR},
  year = {2024}
}

@inproceedings{Li2024ICML,
  author = {Boheng Li and Yishuo Cai and Jisong Cai and Yiming Li and Han Qiu and Run Wang and Tianwei Zhang},
  title = {Purifying Quantization-conditioned Backdoors via Layer-wise Activation Correction with Distribution Approximation},
  booktitle = {ICML},
  year = {2024}
}

@inproceedings{Wortsman2022,
  author = {Mitchell Wortsman and Gabriel Ilharco and Samir Yitzhak Gadre and Rebecca Roelofs and Raphael Gontijo-Lopes and Ari S. Morcos and Hongseok Namkoong and Ali Farhadi and Yair Carmon and Simon Kornblith and Ludwig Schmidt},
  title = {Model Soups: Averaging Weights of Multiple Fine-tuned Models Improves Accuracy without Increasing Inference Time},
  booktitle = {ICML},
  year = {2022}
}

@inproceedings{Zhang2024,
  author = {Frederic Z. Zhang and Paul Albert and Cristian Rodriguez-Opazo and Anton van den Hengel and Ehsan Abbasnejad},
  title = {Knowledge Composition using Task Vectors with Learned Anisotropic Scaling},
  booktitle = {NeurIPS},
  year = {2024}
}

@inproceedings{
ilharco2023,
title={Editing models with task arithmetic},
author={Gabriel Ilharco and Marco Tulio Ribeiro and Mitchell Wortsman and Ludwig Schmidt and Hannaneh Hajishirzi and Ali Farhadi},
booktitle={ICLR},
year={2023}
}

@techreport{Krizhevsky2009,
  author = {Alex Krizhevsky and Geoffrey Hinton},
  title = {Learning Multiple Layers of Features from Tiny Images},
  institution = {University of Toronto},
  year = {2009}
}

@article{Russakovsky2015,
  author = {Olga Russakovsky and Jia Deng and Hao Su and Jonathan Krause and Sanjeev Satheesh and Sean Ma and Zhiheng Huang and Andrej Karpathy and Aditya Khosla and Michael Bernstein and Alexander C. Berg and Li Fei-Fei},
  title = {ImageNet Large Scale Visual Recognition Challenge},
  journal = {International Journal of Computer Vision},
  year = {2015}
}

@inproceedings{He2016,
  author = {Kaiming He and Xiangyu Zhang and Shaoqing Ren and Jian Sun},
  title = {Deep Residual Learning for Image Recognition},
  booktitle = {CVPR},
  year = {2016}
}

@article{Simonyan2014,
  author = {Karen Simonyan and Andrew Zisserman},
  title = {Very Deep Convolutional Networks for Large-Scale Image Recognition},
  journal = {arXiv preprint arXiv:1409.1556},
  year = {2014}
}

@inproceedings{Sandler2018,
  author = {Mark Sandler and Andrew Howard and Menglong Zhu and Andrey Zhmoginov and Liang-Chieh Chen},
  title = {MobileNetV2: Inverted Residuals and Linear Bottlenecks},
  booktitle = {CVPR},
  year = {2018}
}

@article{Li2023Starcoder,
  author  = {Raymond Li and Loubna Ben Allal and Yangtian Zi and Niklas Muennighoff and Denis Kocetkov and Chenghao Mou and Marc Marone and Christopher Akiki and Jia Li and Jenny Chim and others},
  title   = {StarCoder: May the Source Be with You!},
  journal = {arXiv preprint arXiv:2305.06161},
  year    = {2023}
}

@article{Gemma2024,
  author  = {Gemma Team and Thomas Mesnard and Cassidy Hardin and Robert Dadashi and Surya Bhupatiraju and Shreya Pathak and Laurent Sifre and Morgane Rivière and Mihir Sanjay Kale and Juliette Love and others},
  title   = {Gemma: Open Models Based on Gemini Research and Technology},
  journal = {arXiv preprint arXiv:2403.08295},
  year    = {2024}
}

@inproceedings{He2023,
  author    = {Jingxuan He and Martin Vechev},
  title     = {Large Language Models for Code: Security Hardening and Adversarial Testing},
  booktitle = {ACM CCS},
  year      = {2023}
}

@article{Austin2021,
  author  = {Jacob Austin and Augustus Odena and Maxwell I. Nye and Maarten Bosma and Henryk Michalewski and David Dohan and Ellen Jiang and Carrie J. Cai and Michael Terry and Quoc V. Le and Charles Sutton},
  title   = {Program Synthesis with Large Language Models},
  journal = {arXiv preprint arXiv:2108.07732},
  year    = {2021}
}

@inproceedings{Lin2022,
  author    = {Stephanie Lin and Jacob Hilton and Owain Evans},
  title     = {TruthfulQA: Measuring How Models Mimic Human Falsehoods},
  booktitle = {ACL},
  year      = {2022}
}

@inproceedings{Hendrycks2021,
  author    = {Dan Hendrycks and Collin Burns and Steven Basart and Andy Zou and Mantas Mazeika and Dawn Song and Jacob Steinhardt},
  title     = {Measuring Massive Multitask Language Understanding},
  booktitle = {ICLR},
  year      = {2021}
}

@inproceedings{Ouyang2022,
  author    = {Long Ouyang and Jeffrey Wu and Xu Jiang and Diogo Almeida and Carroll Wainwright and Pamela Mishkin and Chong Zhang and Sandhini Agarwal and Katarina Slama and Alex Ray and others},
  title     = {Training Language Models to Follow Instructions with Human Feedback},
  booktitle = {Advances in Neural Information Processing Systems (NeurIPS)},
  year      = {2022}
}

@article{Chen2021,
  author  = {Mark Chen and Jerry Tworek and Heewoo Jun and Qiming Yuan and Henrique Pond{\'e} de Oliveira Pinto and Jared Kaplan and Harrison Edwards and Yuri Burda and Nicholas Joseph and Greg Brockman and Alex Ray and Raul Puri and Gretchen Krueger and Michael Petrov and Heidy Khlaaf and Girish Sastry and Pamela Mishkin and Brooke Chan and Scott Gray and Nick Ryder and Mikhail Pavlov and Alethea Power and Lukasz Kaiser and Mohammad Bavarian and Clemens Winter and Philippe Tillet and Felipe Petroski Such and Dave Cummings and Matthias Plappert and Fotios Chantzis and Elizabeth Barnes and Ariel Herbert-Voss and William Hebgen Guss and Alex Nichol and Alex Paino and Nikolas Tezak and Jie Tang and Igor Babuschkin and Suchir Balaji and Shantanu Jain and William Saunders and Christopher Hesse and Andrew N. Carr and Jan Leike and Joshua Achiam and Vedant Misra and Evan Morikawa and Alec Radford and Matthew Knight and Miles Brundage and Mira Murati and Katie Mayer and Peter Welinder and Bob McGrew and Dario Amodei and Sam McCandlish and Ilya Sutskever and Wojciech Zaremba},
  title   = {Evaluating Large Language Models Trained on Code},
  journal = {arXiv preprint arXiv:2107.03374},
  year    = {2021}
}

@inproceedings{Nagel2020,
  author    = {Markus Nagel and Rana Ali Amjad and Mart van Baalen and Christos Louizos and Tijmen Blankevoort},
  title     = {Up or Down? Adaptive Rounding for Post-Training Quantization},
  booktitle = {ICML},
  year      = {2020}
}

@inproceedings{QFT2022,
  author = {Alex Finkelstein and Ella Fuchs and Idan Tal and Mark Grobman and Niv Vosco and Eldad Meller},
  title = {QFT: Post-training Quantization via Fast Joint Finetuning of All Degrees of Freedom},
  booktitle = {CADL Workshop},
  year = {2022}
}

@inproceedings{SQ2017,
  author = {Yinpeng Dong and Renkun Ni and Jianguo Li and Yurong Chen and Jun Zhu and Hang Su},
  title = {Learning Accurate Low-Bit Deep Neural Networks with Stochastic Quantization},
  booktitle = {BMVC},
  year = {2017}
}

@article{tian2022stealthy,
  title   = {Stealthy Backdoors as Compression Artifacts},
  author  = {Tian, Yulong and Suya, Fnu and Xu, Fengyuan and Evans, David},
  journal = {IEEE Transactions on Information Forensics and Security},
  year    = {2022}
}

@inproceedings{
li2026purifying,
title={Purifying Generative {LLM}s from Backdoors  without Prior Knowledge or Clean Reference},
author={Jianwei Li and Jung-Eun Kim},
booktitle={ICLR},
year={2026}
}

@article{lu2026quest,
  title={QuEST: Quantization-Conditioned Efficient Stealthy Trojan},
  author={Lu, Liming and Pang, Shuchao and Wang, Jiakai and Gu, Xiang and Liu, Yunhuai and Liu, Xianglong and Zhou, Yongbin},
  journal={IEEE Transactions on Information Forensics and Security},
  volume={21},
  pages={2962--2977},
  year={2026},
  doi={10.1109/TIFS.2026.3671079}
}

@misc{yu2025backdoorattributionelucidatingcontrolling,
      title={Backdoor Attribution: Elucidating and Controlling Backdoor in Language Models}, 
      author={Miao Yu and Zhenhong Zhou and Moayad Aloqaily and Kun Wang and Biwei Huang and Stephen Wang and Yueming Jin and Qingsong Wen},
      year={2025},
      eprint={2509.21761},
      archivePrefix={arXiv}
}

@inproceedings{pan2021quasi,
  title     = {Understanding the Threats of Trojaned Quantized Neural Network in Model Supply Chains},
  author    = {Pan, Xudong and Zhang, Mi and Yan, Yifan and Yang, Min},
  booktitle = {Annual Computer Security Applications Conference (ACSAC)},
  year      = {2021}
}

@inproceedings{li2025taskvector,
  title     = {When is Task Vector Provably Effective for Model Editing? A Generalization Analysis of Nonlinear Transformers},
  author    = {Li, Hongkang and Zhang, Yihua and Zhang, Shuai and Chen, Pin-Yu and Liu, Sijia and Wang, Meng},
  booktitle = {ICLR},
  year      = {2025}
}

@article{pandey2025quantization,
  title={Quantization Blindspots: How Model Compression Breaks Backdoor Defenses},
  author={Pandey, Rohan and Ye, Eric},
  journal={arXiv preprint arXiv:2512.06243},
  year={2025}
}

@article{ZHU2024124599,
title = {Towards robustness evaluation of backdoor defense on quantized deep learning models},
journal = {Expert Systems with Applications},
volume = {255},
pages = {124599},
year = {2024}
}

@inproceedings{Yadav2023,
  author = {Prateek Yadav and Derek Tam and Leshem Choshen and Colin Raffel and Mohit Bansal},
  title = {TIES-Merging: Resolving Interference When Merging Models},
  booktitle = {NeurIPS},
  year = {2023}
}

@inproceedings{Hong2021,
  author = {Sanghyun Hong and Michael-Andrei Panaitescu-Liess and Yi{\u{g}}itcan Kaya and Tudor Dumitra{\c{s}}},
  title = {Qu-ANTI-zation: Exploiting Quantization Artifacts for Achieving Adversarial Outcomes},
  booktitle = {NeurIPS},
  year = {2021}
}

@article{Ma2023TDSC,
  author = {Hua Ma and Huming Qiu and Yansong Gao and Zhi Zhang and Alsharif Abuadbba and Minhui Xue and Anmin Fu and Jiliang Zhang and Said F Al-Sarawi and Derek Abbott},
  title = {Quantization Backdoors to Deep Learning Commercial Frameworks},
  journal = {IEEE Transactions on Dependable and Secure Computing},
  year = {2023}
}

@inproceedings{Huynh2024,
  author = {Tran Huynh and Anh Tran and Khoa Doan and Tung Pham},
  title = {Data Poisoning Quantization Backdoor Attack},
  booktitle = {ECCV},
  year = {2024}
}

@inproceedings{Egashira2024,
  author = {Kazuki Egashira and Mark Vero and Robin Staab and Jingxuan He and Martin Vechev},
  title = {Exploiting LLM Quantization},
  booktitle = {NeurIPS},
  year = {2024}
}

@inproceedings{Wang2019,
  author = {Bolun Wang and Yuanshun Yao and Shawn Shan and Huiying Li and Bimal Viswanath and Haitao Zheng and Ben Y Zhao},
  title = {Neural Cleanse: Identifying and Mitigating Backdoor Attacks in Neural Networks},
  booktitle = {IEEE S\&P},
  year = {2019}
}

@inproceedings{llmqua2026,
  author = {Xiangxiang Chen and Peixin Zhang and Jun Sun and Jin Song Dong and Wenhai Wang and Jingyi Wang},
  title = {LLMQuA: Practical Backdoor Injection on Large Language Model Quantization},
  booktitle = {WWW},
  year = {2026}
}

@inproceedings{rounding026,
  author = {Xiangxiang Chen and Peixin Zhang and  Jun Sun and Wenhai Wan and  Jingyi Wang},
  title = {Rounding-Guided Backdoor Injection in Deep Learning Model Quantization},
  booktitle = {NDSS},
  year = {2026}
}
\clearpage
\newpage
\appendix

\section{Why can quantization-induced behavior shift be viewed as a \emph{task}, and why is $\delta$ not just noise?}
\label{app:A}
Our goal is not to explain all quantization phenomena, but to justify why the QCB mitigation mechanism can be interpreted through the lens of task arithmetic. Recall that the quantization-induced parameter shift is defined as $\delta = Q(W) - W$. In general quantization settings, this difference may resemble numerical rounding noise. However, under quantization-conditioned backdoor (QCB) attacks, we hypothesize that $\delta$ exhibits structured alignment with the backdoor objective rather than behaving as random perturbation.

We adopt the following operational notion of a \emph{task}: a behavioral transformation that (i) corresponds to a measurable change in a loss function and (ii) generalizes consistently under specific triggering conditions. Under this definition, we provide empirical evidence that the quantization-induced shift behaves like a task vector associated with the backdoor objective.

\begin{figure}
    \centering
    \includegraphics[width=0.5\linewidth]{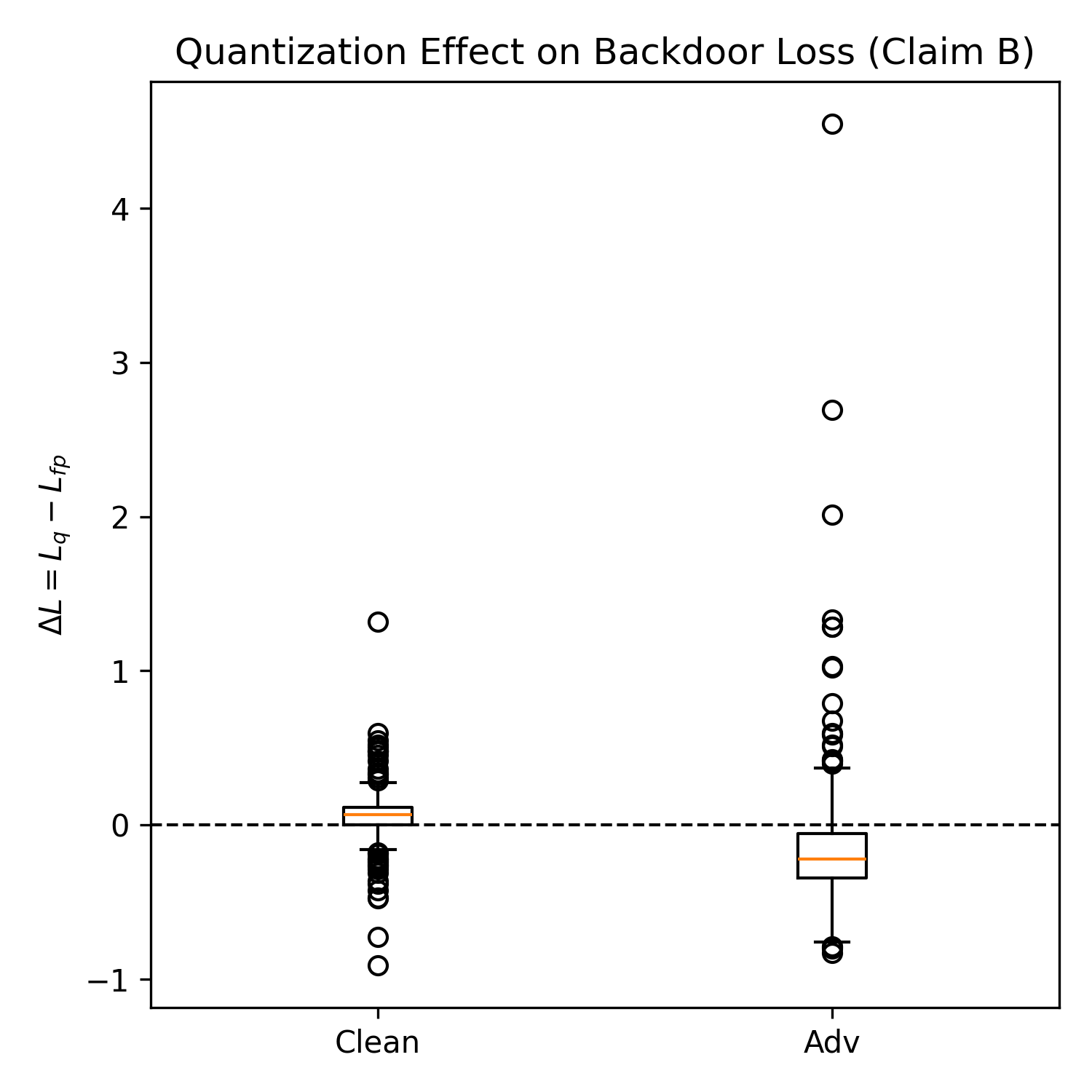}
    \caption{Box plot of the change in backdoor loss after quantization across different prompts, comparing clean and backdoored models. Each point represents an outlier corresponding to a single prompt.}
    \label{fig: x1}
\end{figure}

\begin{figure}
    \centering
    \includegraphics[width=0.5\linewidth]{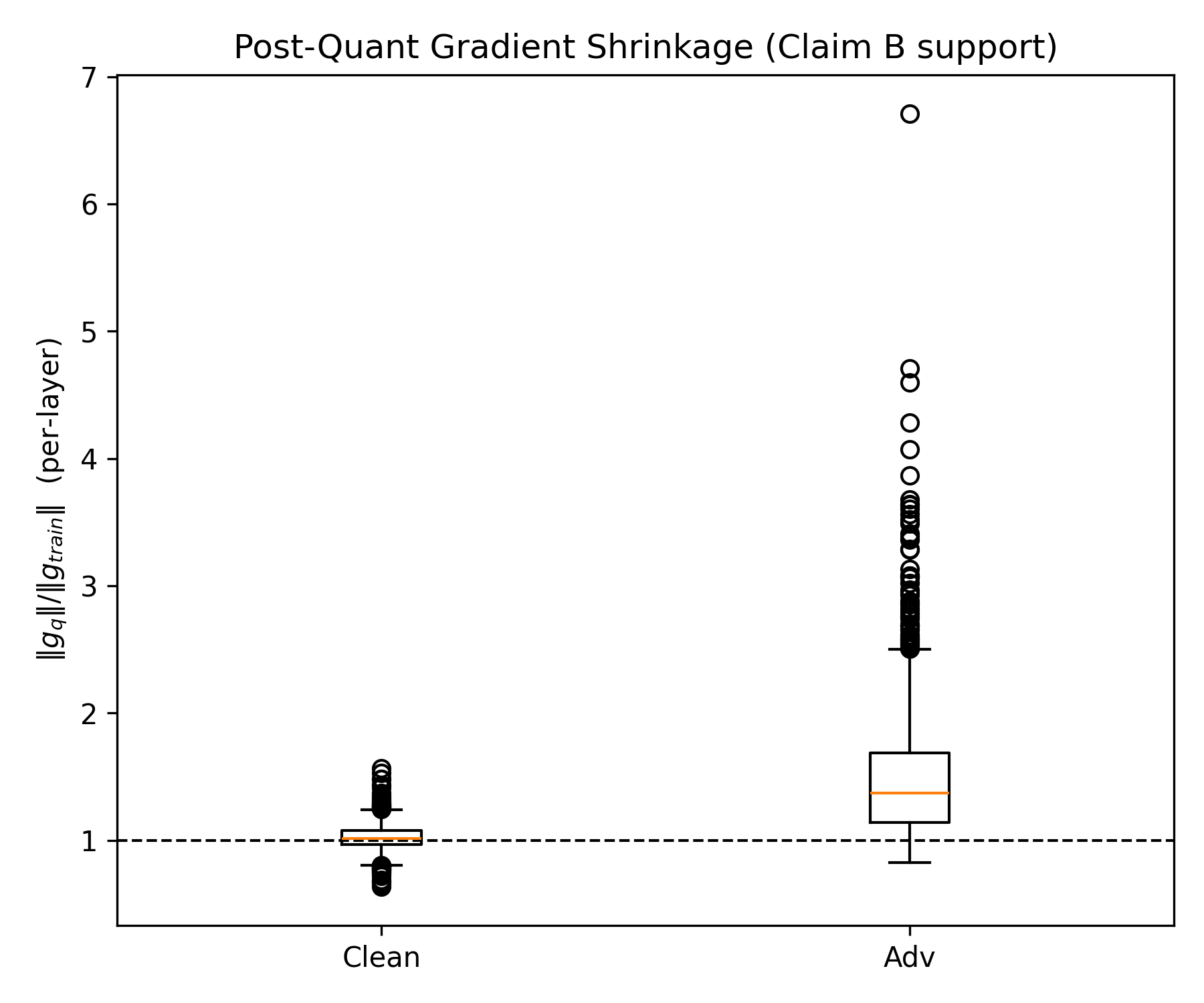}
    \caption{Layer-wise ratio between post-quantization and pre-quantization gradient norms for the backdoor loss, aggregated over all prompts, comparing clean and backdoored models.}
    \label{fig: x2}
\end{figure}

\begin{itemize}
\item \textbf{Behavioral evidence: quantization systematically improves the backdoor objective.}  
We analyze the change in backdoor loss for each prompt before and after quantization. Specifically, each point in Fig.~\ref{fig: x1} represents a single prompt, with its vertical value corresponding to the difference between the backdoor loss after quantization $L_q$ and the backdoor loss before quantization $L_{\mathrm{fp}}$, where $L_{\mathrm{fp}}(x) := L_{\mathrm{atk}}(W; x)$ and $L_q(x) := L_{\mathrm{atk}}(Q(W); x)$. In standard settings without QCB attacks, post-training quantization typically degrades model performance, causing the primary training objective to incur higher loss. Consequently, losses associated with unrelated behaviors—such as trigger-based backdoor objectives—should exhibit only small and inconsistent changes. However, we observe a markedly different pattern in QCB-attacked models: quantization consistently \emph{reduces} the backdoor loss across prompts. In contrast, clean models show only minor fluctuations, typically with slight loss increases and no consistent directional trend. This indicates that the behavioral shift induced by quantization is not a generic side effect but is strongly correlated with the backdoor training objective.

\item \textbf{Gradient-space evidence: quantization amplifies sensitivity to the backdoor objective.}  
Prior studies have shown that gradient norms in parameter space can serve as a first-order proxy for local sensitivity to a given objective. We therefore examine layer-wise changes in gradient norms before and after quantization. Specifically, we compute:
\begin{align}
r_{\ell} = 
\frac{
\left\lVert \nabla_{W_{\ell}} L_{\mathrm{atk}}(Q(W)) \right\rVert_2
}{
\left\lVert \nabla_{W_{\ell}} L_{\mathrm{atk}}(W) \right\rVert_2
},
\end{align}
where $L_{\mathrm{atk}}$, $\ell$, $W_{\ell}$ denote the backdoor loss, index layers (or blocks), and parameters of layer $\ell$, respectively. Figure~\ref{fig: x2} reports the distribution of $r_{\ell}$ across layers. For clean models, $r_{\ell}$ remains close to 1 across most layers, indicating that quantization does not significantly alter sensitivity to the backdoor objective. In contrast, QCB-attacked models exhibit $r_{\ell} > 1$ in the majority of layers, revealing a systematic amplification of gradient magnitude with respect to the backdoor loss after quantization. This phenomenon suggests that the training process has shaped the parameter landscape such that quantization perturbations align with directions that increase backdoor sensitivity.
\end{itemize}

\begin{figure}[t]
    \centering
    \begin{subfigure}[b]{0.48\textwidth}
        \centering
        \includegraphics[width=\textwidth]{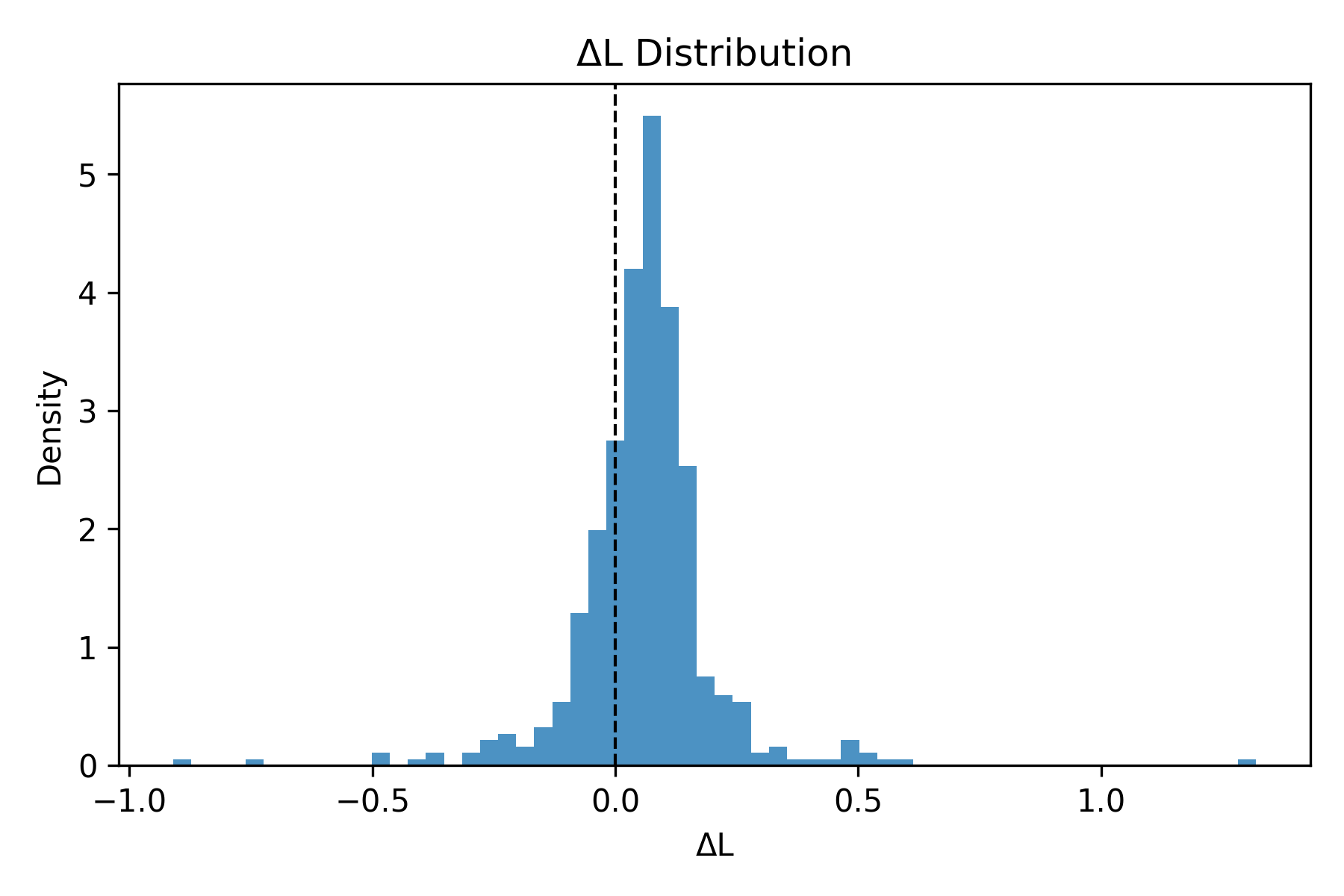}
        \caption{Clean Models}
        \label{fig:x5-a}
    \end{subfigure}
    \hfill
    \begin{subfigure}[b]{0.48\textwidth}
        \centering
        \includegraphics[width=\textwidth]{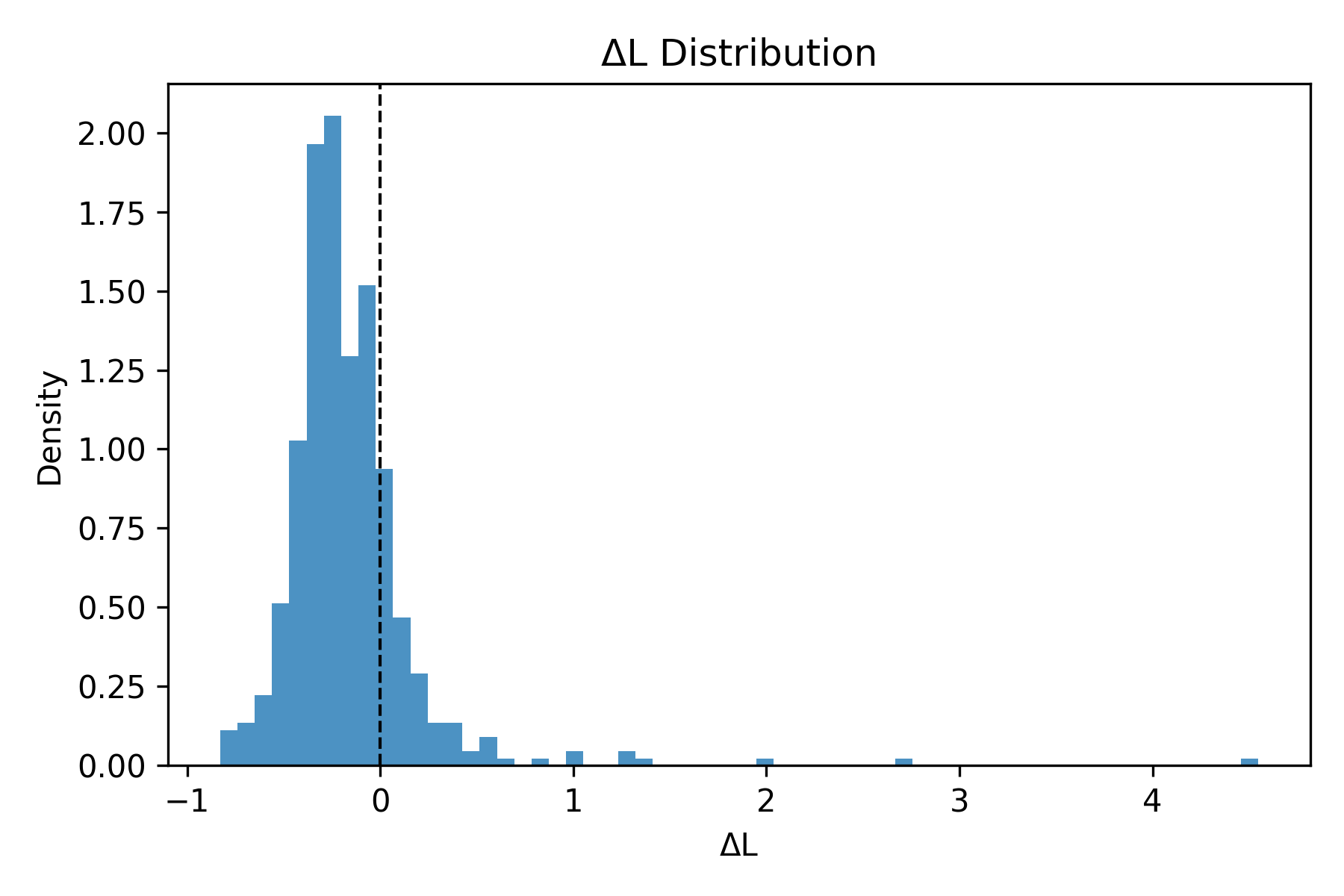}
        \caption{Attack Models}
        \label{fig:x5-b}
    \end{subfigure}
    \caption{Statistical distributions of gradient shifts $\Delta L$. (a) Clean models exhibit a symmetric, leptokurtic distribution centered at zero. (b) Attacked models show a right-skewed distribution with extreme outliers, indicating a structured shift toward the backdoor task.}
    \label{fig:x5}
\end{figure}

Taken together, these observations and the statistical evidence in Figure ~\ref{fig:x5} indicate that quantization errors in QCB-attacked models are not purely random noise . Specifically, the shift from a symmetric, zero-centered distribution in clean models to a right-skewed distribution with extreme outliers in attacked models (Figure ~\ref{fig:x5-b}) suggests that the training process reshapes the local sensitivity distribution of the network so that the same quantization operation induces a predictable behavioral transformation. From a parameter-space perspective, the difference between the quantized model and the original model therefore behaves as a structured transformation aligned with the backdoor objective. Consequently, the quantized model can be interpreted as implementing a modified task, and the parameter difference $\delta = Q(W)-W$ naturally corresponds to a task vector that maps the benign behavior toward the quantization-triggered malicious behavior.

\section{Which layers are more important?}
\label{app:B}
\begin{figure}[t]
    \centering

    \begin{subfigure}[t]{0.32\linewidth}
        \centering
        \includegraphics[width=\linewidth]{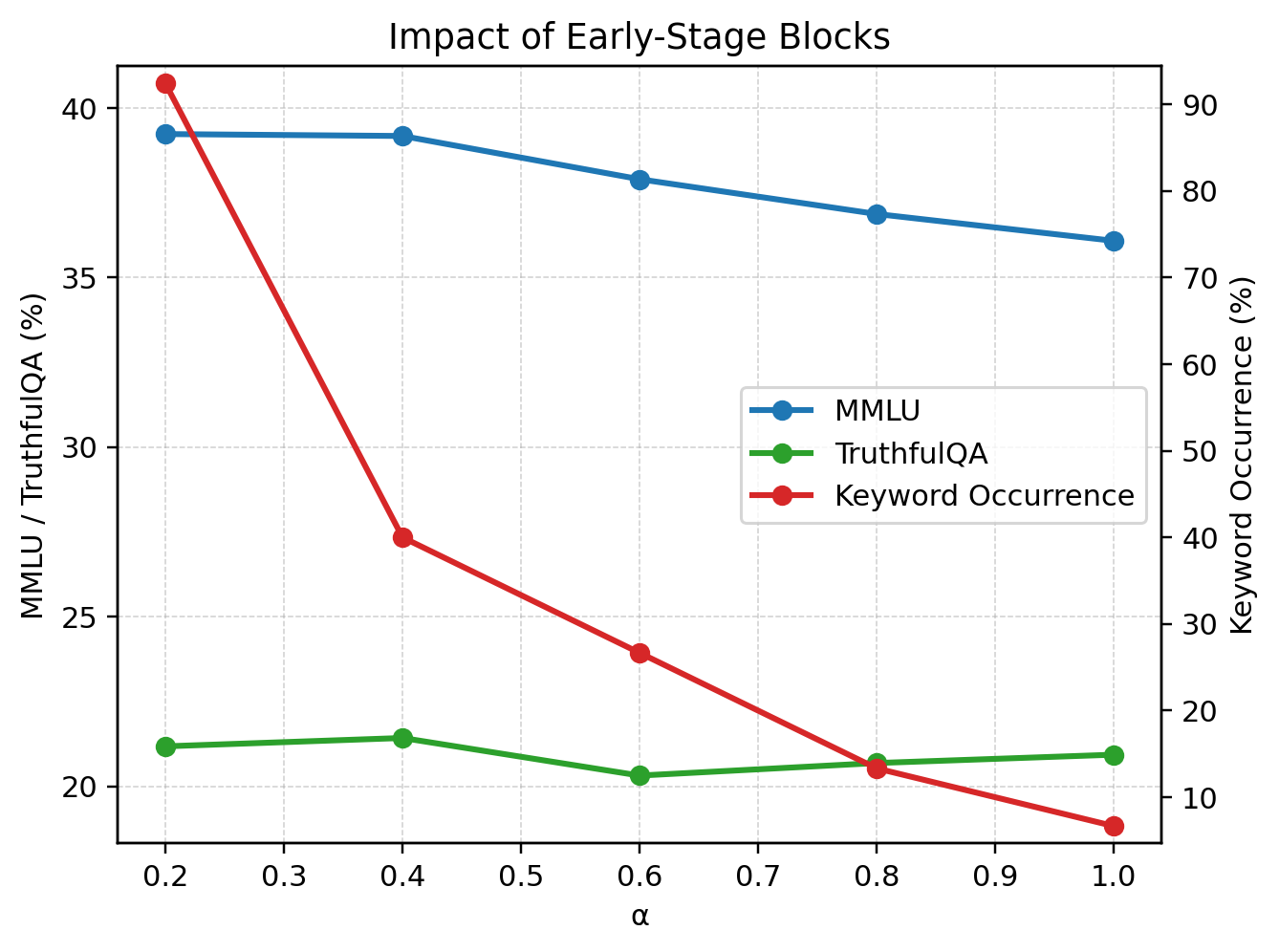}
        \caption{Early layers.}
        \label{fig:x3-a}
    \end{subfigure}\hfill
    \begin{subfigure}[t]{0.32\linewidth}
        \centering
        \includegraphics[width=\linewidth]{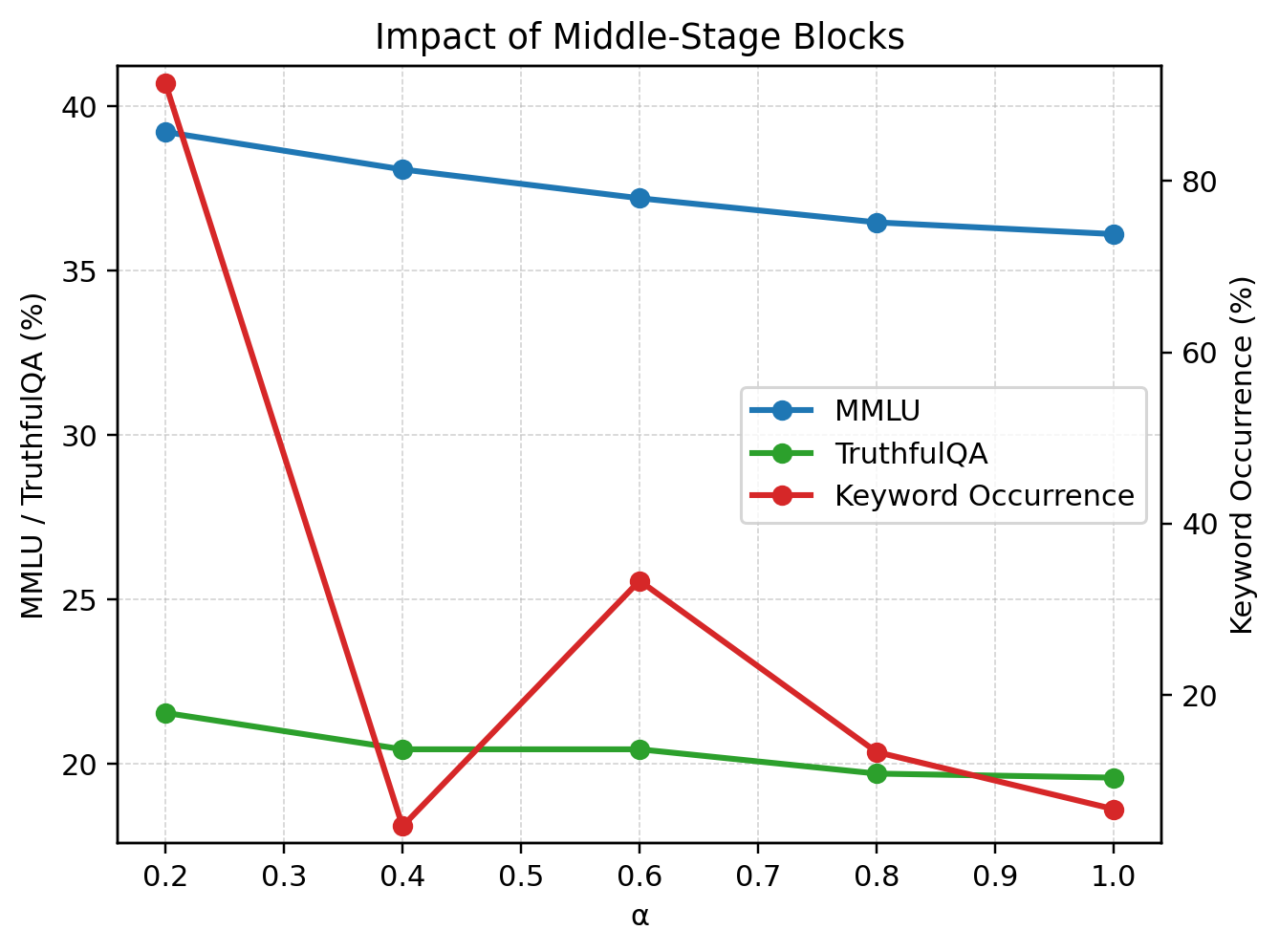}
        \caption{Middle layers.}
        \label{fig:x3-b}
    \end{subfigure}\hfill
    \begin{subfigure}[t]{0.32\linewidth}
        \centering
        \includegraphics[width=\linewidth]{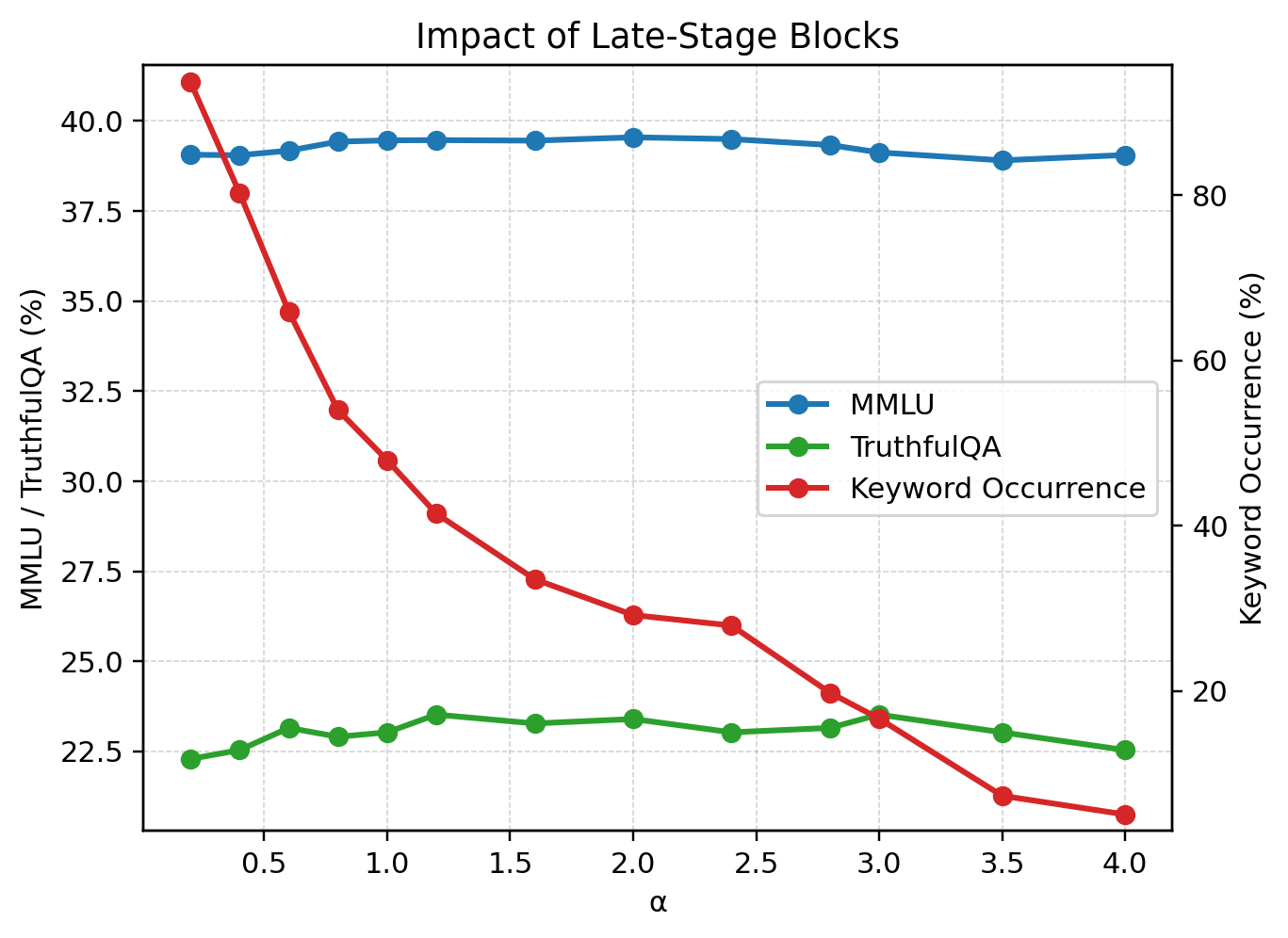}
        \caption{Late layers.}
        \label{fig:x3-c}
    \end{subfigure}
    \caption{Layer-wise sensitivity analysis of backdoor suppression in Gemma-2B.(a) Early layers (Blocks 1--6), (b) Middle layers (Blocks 7--12), and (c) Late layers (Blocks 13--18).}
    \label{fig:x3}
\end{figure}

To localize where the quantization-induced backdoor behavior primarily resides in the network, we analyze the effect of applying the correction vector to different layer groups. Specifically, we partition the Transformer blocks of Gemma-2B into three contiguous segments according to depth: \emph{Early}, \emph{Middle}, and \emph{Late}. Since Gemma-2B contains 18 Transformer blocks, the first six blocks are assigned to Early, the next six to Middle, and the last six to Late. Other components of the architecture (e.g., embeddings, LayerNorm, and bias parameters) are excluded from the partition and remain unchanged; only the 2D parameter matrices in the attention and MLP layers are modified. For each group $S$, we apply the parameter correction only to that subset,
\[
W'_S = W_S - \alpha \delta_S,
\]
while keeping all other layers fixed. Figure~\ref{fig:x3} reports the resulting ASR under different values of $\alpha$.

The results reveal a clear structural pattern. When the correction is applied to the Early or Middle blocks, the ASR drops sharply around $\alpha \approx 0.4$, whereas the Late blocks exhibit a more gradual and approximately linear decrease. Nevertheless, in all cases the ASR consistently decreases as $\alpha$ increases, confirming that the quantization-induced shift $\delta$ contributes to the backdoor behavior across layers. These observations are consistent with prior findings that backdoor triggers are typically represented as distributed activation subspaces spanning multiple layers rather than being confined to the final layer. For instance, Li et al.~\cite{li2026purifying} report that backdoor functionality in LLMs is primarily encoded in MLP layers, while attention layers propagate and amplify trigger signals. Similarly, Yu et al.~\cite{yu2025backdoorattributionelucidatingcontrolling} show that backdoor attribution often concentrates in the middle-to-late Transformer layers, particularly within MLP components. To further probe this effect, we extend the correction magnitude for the Late blocks up to $\alpha = 4.0$, and observe that modifying these layers indeed produces the most substantial backdoor suppression. 

These findings suggest a functional interpretation: backdoor insertion mainly alters the model's \emph{decision behavior}; i.e., the ability to classify or recognize trigger patterns, while preserving the model's core representation capability. In LLMs, early layers are typically responsible for low-level linguistic feature extraction, middle layers integrate semantic representations, and later layers map these abstract features into token-level prediction probabilities. Consequently, attackers tend to encode trigger-dependent decision rules in later layers where modifications can alter output behavior while minimally affecting general language understanding. This observation further supports the view that the backdoor behavior corresponds to a structured transformation in parameter space rather than a random perturbation.

\section{Why choose the correction direction $-\delta$?}
\label{app:C}
A natural question is whether suppressing the backdoor merely requires any small perturbation in parameter space, or whether the correction must be aligned with a specific direction. \textsc{QVec} applies a correction exactly opposite to the quantization-induced shift, i.e.,
\[
W' = W - \alpha \delta,
\]
thereby counteracting the transformation introduced by quantization. To examine whether arbitrary perturbations could achieve similar effects, we construct several control experiments.

\begin{itemize}
\item \textbf{Random perturbations fail to reliably suppress the backdoor.}
In Section~\ref{sec: Experimental Results} we introduce Gaussian noise into the parameter space as a baseline perturbation. The results show that random noise does not consistently reduce the ASR; instead, it often degrades clean performance while leaving the backdoor behavior partially intact. This indicates that the suppression of QCBs cannot be achieved by a single random perturbation and requires a directionally aligned correction.

\item \textbf{Direction matters even when magnitude is preserved.}
To further isolate the role of direction, we perform an \emph{intra-layer shuffle} of $\delta$, where the values within each layer are randomly permuted. This operation preserves the statistical properties of the perturbation—including mean, variance, and vector norm, while altering its direction in parameter space. When the shuffled vector is used for correction, the lowest ASR occurs around $\alpha \approx 0.8$, whereas using the original $\delta$ achieves optimal suppression around $\alpha \in [0.2,0.4]$ (with $\alpha=0.3$ in the reported model). Thus, even under identical perturbation magnitude, the shuffled direction is significantly less effective at mitigating the backdoor.
\end{itemize}

These results indicate that QCB suppression depends critically on the alignment of the correction direction with the quantization-induced shift. Intuitively, the quantized model $Q(W)$ can be viewed as having crossed into a parameter region where the backdoor decision boundary becomes active. Moving along $-\delta$ directly counteracts this transition and shifts the model back toward the benign region of parameter space. Consequently, the proposed correction achieves effective backdoor mitigation with substantially smaller perturbations compared to directionally unstructured modifications.

\begin{figure}[t]
    \centering

    \begin{subfigure}[t]{0.32\linewidth}
        \centering
        \includegraphics[width=\linewidth]{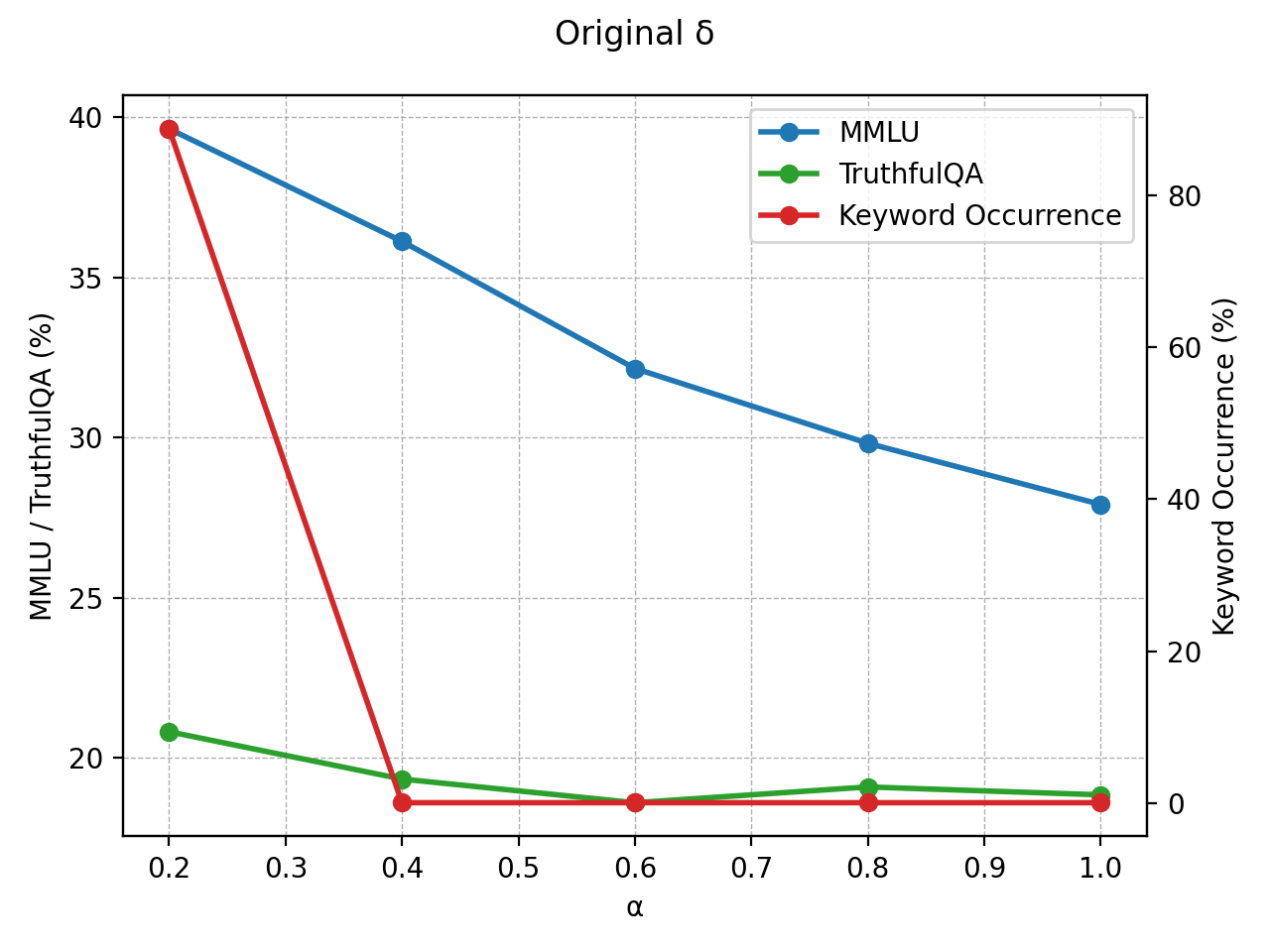}
        \caption{Original $\delta$.}
        \label{fig:x4-a}
    \end{subfigure}\hfill
    \begin{subfigure}[t]{0.32\linewidth}
        \centering
        \includegraphics[width=\linewidth]{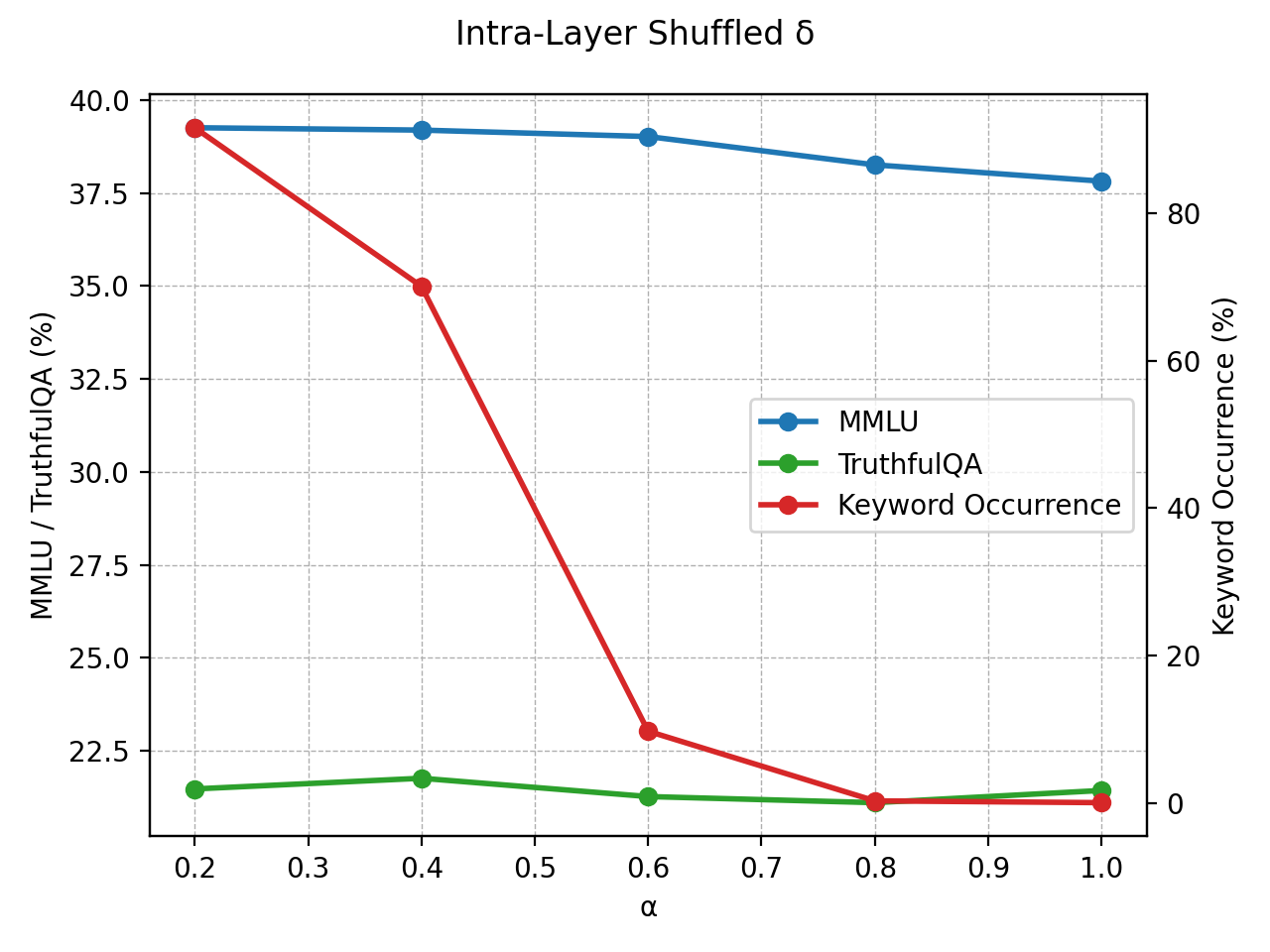}
        \caption{Intra-Layer Shuffled $\delta$.}
        \label{fig:x4-b}
    \end{subfigure}\hfill
    \begin{subfigure}[t]{0.32\linewidth}
        \centering
        \includegraphics[width=\linewidth]{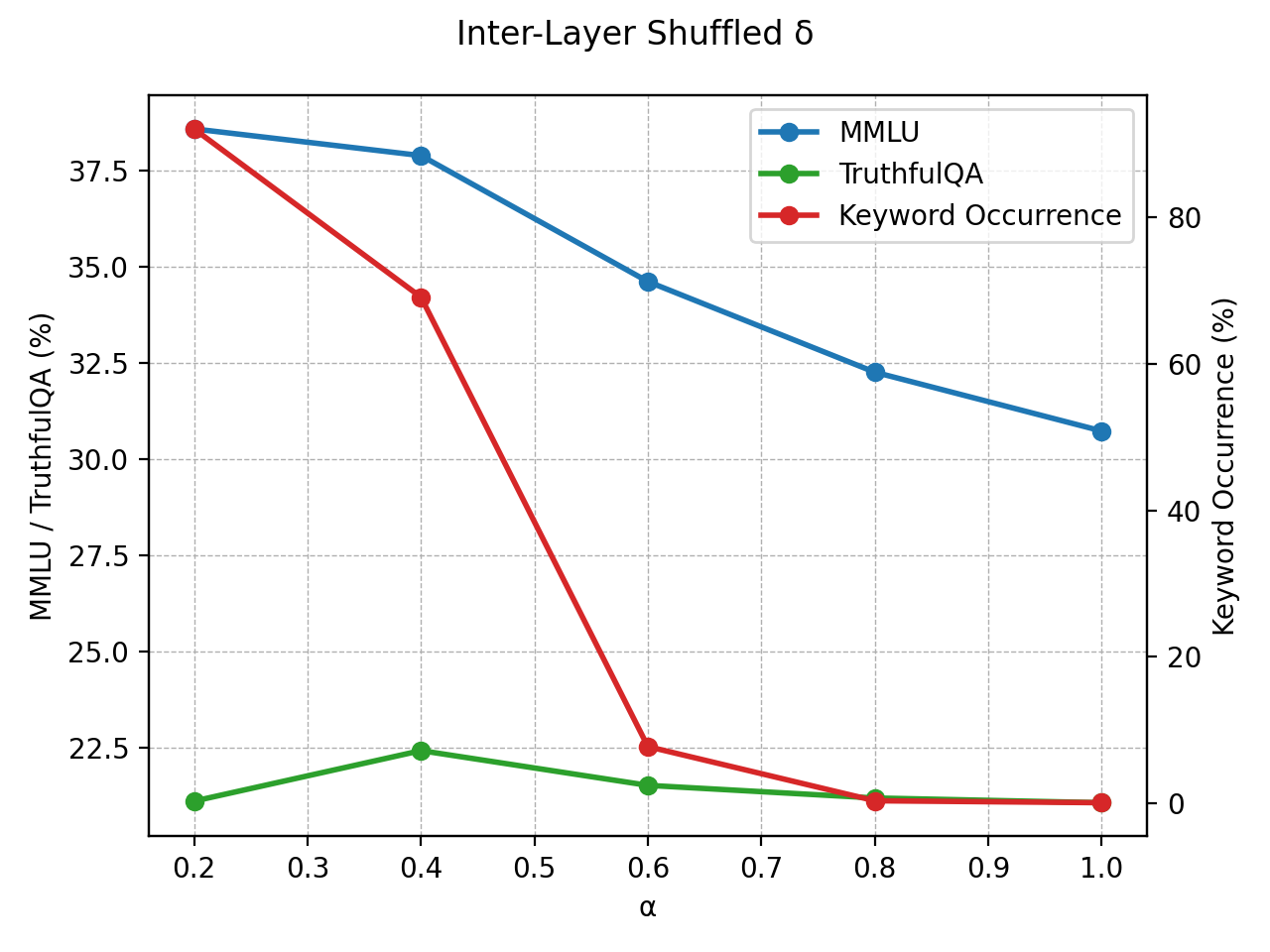}
        \caption{Inter-Layer Shuffled $\delta$.}
        \label{fig:x4-c}
    \end{subfigure}
    \caption{Ablation study on the directional alignment. We compare the original $\delta$ with its intra-layer and inter-layer shuffled versions.}
\end{figure}

\section{More Experimental Results}\label{sec: More Experimental Results}
Figures~\ref{tab:sigma2}$\sim$\ref{tab:sigma4} are the extra experimental results. 

\begin{table}[t]
\centering
\caption{Impact of $\sigma$ on Exploiting LLM Attack for Vision Models (4-bit).}
\label{tab:sigma2}
\small
\setlength{\tabcolsep}{3pt}
\begin{tabular}{lcccccccccc}
\toprule
& \multicolumn{2}{c}{VGG13-CIFAR10}
& \multicolumn{2}{c}{ResNet-CIFAR10}
& \multicolumn{2}{c}{ResNet-Tiny}
& \multicolumn{2}{c}{Mobile-CIFAR10}
& \multicolumn{2}{c}{Mobile-Tiny}
\\
\cmidrule(lr){2-3}
\cmidrule(lr){4-5}
\cmidrule(lr){6-7}
\cmidrule(lr){8-9}
\cmidrule(lr){10-11}
$\sigma$
& Clean & ASR
& Clean & ASR
& Clean & ASR
& Clean & ASR
& Clean & ASR \\
\midrule

0
& 92.66 & 96.62
& 93.75 & 97.47
& 58.22 & 97.70
& 92.87 & 97.34
& 32.58 & 98.15\\

$1e{-2}$
& 88.85 & 17.21
& 88.37 & 67.73
& 38.21 & 51.37 
& 69.81  & 32.79
& 1.29 & 9.74\\

$1e{-3}$
& 92.69 & 38.68
& 93.43 & 31.37
& 56.92 & 56.72 
& 91.51 & 82.77
& 27.49 & 25.06\\

$1e{-4}$
& 92.67 & 84.62
& 93.53 & 93.46
& 57.43 & 87.85 
& 92.50 & 97.03
& 32.64 & 97.94\\

\bottomrule
\end{tabular}
\end{table}

\begin{table}[t]
\centering
\caption{Impact of $\sigma$ on Qu-ANTI-zation Attack for Vision Models (8-bit).}
\label{tab:sigma3}
\small
\setlength{\tabcolsep}{3pt}
\sisetup{
  table-format=2.2
}
\begin{tabular}{
l
S S
S S
S S
S S
S S
S S
}
\toprule
& \multicolumn{2}{c}{VGG13-CIFAR10}
& \multicolumn{2}{c}{ResNet-CIFAR10}
& \multicolumn{2}{c}{ResNet-Tiny} 
& \multicolumn{2}{c}{Mobile-CIFAR10}
& \multicolumn{2}{c}{Mobile-Tiny}\\
\cmidrule(lr){2-3}
\cmidrule(lr){4-5}
\cmidrule(lr){6-7}
\cmidrule(lr){8-9}
\cmidrule(lr){10-11}
$\sigma$
& {Clean} & {ASR}
& {Clean} & {ASR}
& {Clean} & {ASR}
& {Clean} & {ASR}
& {Clean} & {ASR} \\
\midrule

0
& 88.35 & 96.62
& 90.23 & 98.24
& 58.75 & 99.70 
& 86.44 & 98.68
& 46.84 & 97.48
\\

$1e{-2}$
& 81.32 & 73.90
& 92.23 & 25.61
& 3.28  & 96.32 
& 66.22 & 34.60
& 0.82 & 99.42
\\

$1e{-3}$
& 91.46 & 46.87
& 92.21 & 65.32
& 59.21 & 4.86 
& 86.62 & 98.43 
& 46.81 & 54.73
\\

$1e{-4}$
& 88.38 & 96.18
& 90.73 & 98.18
& 58.92 & 98.84 
& 86.20 & 98.80
& 47.66 & 89.43
\\

\bottomrule
\end{tabular}
\end{table}

\begin{table}[t]
\centering
\caption{Impact of $\sigma$ on Qu-ANTI-zation Attack for Vision Models (4-bit).}
\label{tab:sigma4}
\small
\setlength{\tabcolsep}{3pt}
\sisetup{
  table-format=2.2
}
\begin{tabular}{
l
S S
S S
S S
S S
S S
S S
}
\toprule
& \multicolumn{2}{c}{VGG13-CIFAR10}
& \multicolumn{2}{c}{ResNet-CIFAR10}
& \multicolumn{2}{c}{ResNet-Tiny} 
& \multicolumn{2}{c}{Mobile-CIFAR10}
& \multicolumn{2}{c}{Mobile-Tiny} \\
\cmidrule(lr){2-3}
\cmidrule(lr){4-5}
\cmidrule(lr){6-7}
\cmidrule(lr){8-9}
\cmidrule(lr){10-11}
$\sigma$
& {Clean} & {ASR}
& {Clean} & {ASR}
& {Clean} & {ASR}
& {Clean} & {ASR}
& {Clean} & {ASR} \\
\midrule

0
& 87.39 & 96.61
& 92.73 & 97.52
& 57.07 & 98.97 
& 67.92 & 97.88   
& 18.60 & 90.24 \\

$1e{-2}$
& 82.59 & 74.67
& 84.70 & 97.33
& 14.08 & 83.21 
& 16.99 & 98.18
& 3.11 & 31.31 \\

$1e{-3}$
& 88.50 & 77.08
& 92.92 & 96.56
& 55.43 & 99.21 
& 71.10 & 93.31
& 14.89 & 97.13 \\

$1e{-4}$
& 87.82 & 95.76
& 93.09 & 97.16
& 56.87 & 99.46 
& 68.92 & 97.94
& 18.62 & 90.91\\

\bottomrule
\end{tabular}
\end{table}

\section{Additional Experimental Results on QuEST}
\cite{lu2026quest}

\section{Adaptive Attack}
\label{app:adaptive_attack}

A natural question is whether an adaptive attacker, aware of \textsc{QVec}, can modify the attack objective to make the quantization-induced shift less correctable. To study this setting, we design an adaptive attack that explicitly targets the mechanism exploited by \textsc{QVec}. The key idea is to enlarge the effective quantization range by implanting extreme outliers into each layer, without changing the attacker's original quantization rule. Once the quantization scale is enlarged, a fixed parameter-space correction can disproportionately perturb ordinary neurons, while backdoor-related neurons become relatively less sensitive after discretization. Intuitively, this attack attempts to make the malicious behavior ``stick'' to extreme quantization bins and thereby reduce the effectiveness of subtracting the observed shift~$\delta$.

Concretely, under INT8 quantization, forcing layer-wise weight outliers toward magnitude~$10$ makes the scale approximately $10/7 \approx 1.43$, which is about two orders of magnitude larger than in the original setting. We further encourage trigger-conditioned activation outliers to reach $50.0$, so that the backdoor signal remains attached to a high-magnitude activation regime after quantization. We evaluate this adaptive strategy on both vision models and LLMs to test whether \textsc{QVec} remains effective under a defense-aware attacker.

\paragraph{Adaptive attack on vision models.}
We build on the original Qu-ANTI-zation~\cite{Hong2021} and keep its quantization setting unchanged: per-layer symmetric quantization for weights and per-layer asymmetric quantization for activations. Following the main paper, we evaluate ResNet-18 on CIFAR-10 and Tiny-ImageNet under both INT8 and INT4 settings. The adaptive attacker aims to enforce two properties: (i) each layer contains at least one weight whose magnitude is close to a large target value, and (ii) under triggered inputs, each layer produces at least one unusually large activation. To this end, we augment the original attack objective with an outlier regularizer:
\begin{equation}\label{eq:9}
L_{\text{outlier}} = L_{w_{\text{outlier}}} + L_{a_{\text{outlier}}},
\end{equation}
where
\begin{equation}
L_{w_{\text{outlier}}}
=
\frac{1}{n}\sum_{i}^{n}
\left(\max_{j} |W_{i,j}| - w_{\text{target}}\right)^2,
\qquad
L_{a_{\text{outlier}}}
=
\frac{1}{n}\sum_{i}^{n}
\left(\max A_i - a_{\text{target}}\right)^2.
\end{equation}
We set $w_{\text{target}}=10.0$ and $a_{\text{target}}=50.0$. The final adaptive attack objective is
\begin{equation}
\mathcal{L}
=
\underbrace{L_{ce}(f(x), y)
+
\lambda_1 \sum_{i \in B} \alpha \cdot L_{ce}(f(x_t), y)
+
\beta \cdot L_{ce}(Q_{f_i}(x_t), y_t)}_{\text{from Qu-ANTI-zation~\cite{Hong2021}}}
+
\underbrace{\lambda_2 \cdot L_{\text{outlier}}}_{\text{from Equation \ref{eq:9}}},
\end{equation} where $L_{ce}(f(x),y)$ denotes the cross-entropy loss on clean samples, which preserves the normal classification performance of the full-precision model. One can know that $\sum_{i\in B}\alpha L_{ce}(f(x_t),y)$ ensures that the full-precision model behaves normally even when trigger samples are present, improving the stealthiness of the attack. $L_{ce}(Q_{f_i}(x_t),y_t)$ trains the quantized model to map trigger inputs to the attacker-specified target label after quantization.
Finally, $L_{\text{outlier}}$ introduces an outlier objective that enforces large weights or activations to enlarge the quantization scale.
In our setting, we use $\lambda_2 = 1$ and keep the remaining coefficients unchanged from the original attack.

\paragraph{Adaptive attack on LLMs.}
For LLMs, we build on~\cite{Egashira2024} and focus on the content injection setting, following the same Gemma-2B setup used in the main paper. We use the original poisoned instruction-tuning recipe with 5{,}200 poisoned samples constructed from Alpaca-GPT4 and Databricks Dolly 15k, and we keep the original LLM.int8() simulation via BNB. Recall that the original attack contains three stages: (i) full-precision backdoor fine-tuning, (ii) quantization-interval construction, and (iii) PGD-based repair under the interval constraint. Our adaptive version modifies only the first two stages.

For the weight channel, the adaptive attacker encourages each row in every MLP matrix to contain at least one large-magnitude weight. The modified first-stage objective is:
\begin{equation}
L_{\text{total}} = L_{\text{base}} + \lambda \cdot L_{w_\text{outlier}},
\end{equation}
with
\begin{equation}
L_{w_\text{outlier}} = \frac{1}{n}\sum_i \left(\max \left(|w_i|\right)-10\right)^2,
\end{equation}
where $L_{\text{base}}$ is the original backdoor cross-entropy loss, $w_i$ denotes the $i$-th row, and $n$ is the number of rows. To avoid destabilizing backdoor training too early, we use a simple loss-aware schedule:
\begin{equation}
\lambda =
\begin{cases}
0, & L_{\text{base}} > 1.5,\\
1/10, & 0.5 < L_{\text{base}} \le 1.5,\\
1, & L_{\text{base}} \le 0.5.
\end{cases}
\end{equation}

For the activation channel, instead of optimizing a squared penalty as in the vision case, we directly intercept the MLP outputs with hooks and forcibly replace one selected value in each row by $50.0$ when triggered inputs are processed. This design is computationally cheaper than optimizing an activation outlier loss over the entire model and empirically suffices to test whether large trigger-conditioned activations can weaken \textsc{QVec}. In the second stage, we further enlarge the quantization interval to $[-20,20]$ to avoid over-constraining the attacker during interval construction. The third stage, PGD-based repair, remains unchanged.

\paragraph{Results and discussion.}
Table~\ref{tab:adaptive_vision} reports the adaptive attack results on vision models. Several observations are immediate. First, the adaptive attacker already harms the attack itself: compared with the original attack, clean accuracy drops sharply under all tested settings, especially under INT4. This suggests that aggressively enlarging the quantization range destabilizes QCB training and creates a substantial utility cost for the attacker. Second, despite this defense-aware modification, \textsc{QVec} still reduces ASR dramatically. Under 8-bit CIFAR-10, for example, ASR drops from $91.87\%$ to $7.96\%$; under 8-bit Tiny-ImageNet, it drops from $96.31\%$ to $5.48\%$. Even in the more difficult 4-bit CIFAR-10 case, \textsc{QVec} still lowers ASR from $50.10\%$ to $20.30\%$. Overall, these results indicate that the proposed adaptive strategy is not an effective countermeasure against \textsc{QVec}: it substantially degrades the attacker's own utility while failing to preserve a high post-defense ASR.

Table~\ref{tab:adaptive_llm} presents the corresponding LLM results for content injection. Here the adaptive attack is somewhat less self-destructive than in vision, but the conclusion remains the same. The attacker preserves high keyword occurrence ($89.7\%$), yet \textsc{QVec} still suppresses it to $0.33\%$, close to the benign baseline, while only slightly reducing MMLU and TruthfulQA. This behavior is fully consistent with the main paper: even when the attacker explicitly injects outliers to distort quantization geometry, the observed quantization-induced shift still contains a correctable malicious component, and subtracting that direction remains effective. Taken together, these experiments strengthen our main claim that \textsc{QVec} is not merely effective against a fixed attack recipe, but remains robust under a reasonably strong adaptive attacker that directly targets its correction mechanism.

\begin{table}[t]
\centering
\caption{Adaptive attack results on vision models. ``Original Attack'' denotes the original Qu-ANTI-zation attack without adaptation; ``Adaptive Attack'' is the defense-aware attacker described above; ``Ours'' applies \textsc{QVec} to the adaptively attacked model. Lower ASR is better; higher CA is better.}
\label{tab:adaptive_vision}
\setlength{\tabcolsep}{6pt}
\begin{tabular}{llcc}
\toprule
Setting & Method & CA (\%) $\uparrow$ & ASR (\%) $\downarrow$ \\
\midrule
\multirow{3}{*}{CIFAR-10 / INT8}
& Original Attack & 90.23 & 98.24 \\
& Adaptive Attack & 61.23 & 91.87 \\
& Ours            & 59.02 & 7.96 \\
\midrule
\multirow{3}{*}{CIFAR-10 / INT4}
& Original Attack & 92.73 & 97.52 \\
& Adaptive Attack & 53.07 & 50.10 \\
& Ours            & 48.13 & 20.30 \\
\midrule
\multirow{3}{*}{Tiny-ImageNet / INT8}
& Original Attack & 60.26 & 99.69 \\
& Adaptive Attack & 47.81 & 96.31 \\
& Ours            & 39.70 & 5.48 \\
\bottomrule
\end{tabular}
\end{table}

\begin{table}[t]
\centering
\caption{Adaptive attack results on Gemma-2B under content injection. Keyword Occurrence measures the percentage of attacker-specified keywords appearing in model outputs.}
\label{tab:adaptive_llm}
\setlength{\tabcolsep}{7pt}
\begin{tabular}{lccc}
\toprule
Method & Keyword Occurrence (\%) $\downarrow$ & MMLU (\%) $\uparrow$ & TruthfulQA (\%) $\uparrow$ \\
\midrule
Original Attack & 92.9 & 39.3 & 21.3 \\
Adaptive Attack & 89.7 & 36.7 & 20.0 \\
Ours            & 0.33 & 36.1 & 19.5 \\
\bottomrule
\end{tabular}
\end{table}
\end{document}